\documentclass[a4paper,preprint,superscriptaddress,nofootinbib]{revtex4-1}
\usepackage{graphicx}[dvipdfmx]
\usepackage{bm}
\usepackage{amssymb}
\usepackage{amsmath}
\usepackage{color}
\usepackage{cancel}
\usepackage{comment}
\usepackage{here}
\usepackage{multirow}
\usepackage{url}

\begin{document} 

\title{New benchmark scenarios of electroweak baryogenesis \\ in aligned two Higgs double models}

\author{Kazuki Enomoto}
\email{k\_enomoto@hep-th.phys.s.u-tokyo.ac.jp}
\affiliation{Department of Physics, University of Tokyo, Bunkyo-ku, Tokyo 113-0033, Japan}

\author{Shinya Kanemura}
\email{kanemu@het.phys.sci.osaka-u.ac.jp }
\affiliation{Department of Physics, Osaka University, Toyonaka, Osaka 560-0043, Japan}

\author{Yushi Mura}
\email{y\_mura@het.phys.sci.osaka-u.ac.jp }
\affiliation{Department of Physics, Osaka University, Toyonaka, Osaka 560-0043, Japan}


\preprint{OU-HET-1147}

\begin{abstract}
We discuss electroweak baryogenesis in aligned two Higgs doublet models.
It is known that in this model the severe constraint from the experimental results for the electron electric dipole moment can be avoided by destructive interference among CP-violating effects in the Higgs sector.
In our previous work, we showed that the observed baryon number in the Universe can be explained without contradicting current available data in a specific scenario in the same model.
We here first discuss details of the evaluation of baryon number based on the WKB method taking into account all order of the wall velocity.
We then investigate parameter spaces which are allowed under the current available data from collider, flavor and electric dipole moment experiments simultaneously.
We find several benchmark scenarios which can explain baryon asymmetry of the Universe.
We also discuss how we can test these benchmark scenarios at future collider experiments, various flavor experiments and gravitational wave observations.
\end{abstract}

\maketitle

\section{Introduction}

Baryon Asymmetry of the Universe (BAU) is one of the big mysteries of particle physics and cosmology.
By the observation based on the Big-Bang Nucleosynthesis (BBN), the ratio of the (anti) baryon number density $n_B$ ($n_{\overline{B}}$) and the entropy density $s$ is given by~\cite{ParticleDataGroup:2020ssz}
\begin{equation}
\eta_B^{\mathrm{BBN}} = \frac{n_B -n_{\overline{B}}}{s} = (8.2 \mathrm{-} 9.2) \times 10^{-11} ~ \mathrm{at~95 \%~C.L.} 
\label{eq:OBSBAU}
\end{equation}
The CMB observation also gives a consistent results~\cite{Planck:2018vyg}.
The Standard Model (SM) cannot explain the BAU.
Baryogenesis is a promising idea that the BAU was produced by some mechanisms from an incident baryon symmetric world in the early Universe.
In viable models to realize baryogenesis, the Sakharov conditions~\cite{Sakharov:1967dj} have to be satisfied, (1) existence of the baryon number violating interactions, (2) both C and CP being violated, (3) departure from thermal equilibrium.
Scenarios for baryogenesis ever proposed satisfy these conditions by various mechanisms such as GUT baryogenesis~\cite{GUT_Baryogenesis}, Affleck--Dine mechanism~\cite{Affleck:1984fy}, Electroweak Baryogenesis (EWBG)~\cite{Kuzmin:1985mm}, Leptogenesis~\cite{Fukugita:1986hr}, et cetera.

Among these scenarios for baryogenesis, a special interest is in EWBG, where baryon number is generated by the physics at the electroweak scale, depending on the physics of non-minimal Higgs sectors.
Therefore, such a model is testable at experiments.

Although a Higgs boson was discovered, the structure of the Higgs sector remains unknown.
Possibilities of various non-minimal Higgs models are important to be considered especially in connection with new physics beyond the SM.
As the nature of the Higgs sector will be thoroughly explored in the near future, the models of EWBG should also be intensively studied in this timing.

In models of EWBG, the Sakharov conditions are satisfied as follows: (1) baryon number non conservation is realized by sphaleron transition at high temperatures, (2) C is violated because of an electroweak gauge theory, and CP can be violated additionally by interactions of the non-minimal Higgs sector, (3) departure from thermal equilibrium is realized by the strongly first order electroweak phase transition.
Notice that in the case of the SM, both (2) and (3) cannot be satisfied~\cite{Huet:1994jb,EWPT_SM} in a compatible way with the current data, so that extension of the SM is necessary for successful EWBG.

Two Higgs Doublet Models (THDMs) are one of the minimal extensions of the Higgs sector, which can provide useful property required for EWBG.
Hence, EWBG has been investigated in THDMs by many authors for more than three decades~\cite{Turok:1990zg, Cline:1995dg, Fromme:2006cm, Cline:2011mm, Tulin:2011wi, Liu:2011jh, Ahmadvand:2013sna, Chiang:2016vgf, Guo:2016ixx, Fuyuto_Senaha, Dorsch:2016nrg, Modak_Senaha, Basler:2021kgq, Enomoto:2021dkl}.
The first order phase transition may be realized by non-decoupling quantum effects of additional bosons in the effective potential at finite temperatures, which can make the electroweak phase transition to be strongly first order~\cite{Turok:1991uc, Anderson:1991zb, Land:1992sm, Hammerschmitt:1994fn, Cline:1996mga, Laine:2000rm, Blinov:2015sna, Inoue:2015pza, Basler:2016obg, Andersen:2017ika}.
It is well known that the same non-decoupling effects can also predict large deviation from the SM value in the triple Higgs boson coupling at zero temperature, by which the first order phase transition can be tested at future collider experiments \cite{Kanemura:2002vm, Kanemura:2004ch, Kanemura:2004mg, Braathen_Kanemura}.
In 2006, Fromme, Huber and Seniuch~\cite{Fromme:2006cm} had first evaluated the BAU in the top quark transport scenario by the WKB method~\cite{Joyce:1994fu, Joyce:1994zn, Cline:2000nw, Fromme:2006wx, Cline:2020jre} in the THDM with a softly broken $Z_2$ symmetry to avoid Flavor Changing Neutral Currents (FCNCs).
After the Higgs boson discovery in 2012, the benchmark scenario they proposed has not been consistent any more against the current severe constraints on the electron Electric Dipole Moment (EDM)~\cite{ACME:2018yjb} and also the constraint from ATLAS~\cite{ATLAS:2019nkf} and CMS~\cite{CMS:2018uag} results.
Therefore, a new scenario for EWBG has been required to be compatible with the current data.
In the model  with a singlet scalar extension, the CP-violating phase can be introduced in the scalar sector with avoiding the EDM constraint, and non-thermal tree level effects or thermal one loop effects of singlet scalar bosons can make a potential barrier between the symmetric vacuum and the broken vacuum for the strongly first order phase transition~\cite{Espinosa:2011eu, Cline:2012hg, Grzadkowski:2018nbc, Cline:2021iff}.

On the other hand, in THDMs for EWBG, CP violation in the Higgs sector has to be compatible with the EDM data by some cancelation mechanisms~\cite{Fuyuto_Senaha, Cheung:2020ugr, Kanemura:2020ibp, Enomoto:2021dkl}.
For example, in ref.~\cite{Kanemura:2020ibp}, a new scenario has been discussed for the THDM in which significant CP-violating phases are included while current electron EDM data are satisfied due to destructive interference of multiple CP-violating phases in the Higgs sector.
In this model, coupling constants of the Higgs boson with the mass of 125 GeV coincide with those in the SM at the tree level by assuming that there are no mixings among neutral Higgs bosons.
In order to avoid the constraint from FCNCs an alignment is imposed in the Yukawa interactions~\cite{Pich:2009sp}.
Collider phenomenology of the model has been investigated in refs.~\cite{Kanemura:2021atq, Kanemura:2021dez}.
In ref.~\cite{Enomoto:2021dkl}, it was shown that the observed BAU can be explained by EWBG in this model under current available experimental constraints.
It was also found that this model has rich phenomenological predictions which can be tested at future experiments.

In the present paper, we discuss benchmark points and phenomenological consequences of this model for EWBG.
We first discuss details of the evaluation of baryon number based on the WKB method with taking into account all order of the wall velocity~\cite{Cline:2020jre}.
We also show all the formulae used for our analyses.
Second, we investigate parameter spaces which are simultaneously allowed under the current available data from collider~\cite{ATLAS:2019nkf, CMS:2018uag, ATLAS:2020zms, CMS:2019pzc, ATLAS:2018rvc, ALEPH:2013htx, ATLAS:2018gfm, CMS:2019bfg, ATLAS:2021upq, ATLAS:2021yyr}, flavor~\cite{HFLAV:2019otj, Haller:2018nnx, BSG_BaBar, BSG_Belle, CLEO:2001gsa, ATLAS:2018cur, CMS:2019bbr, LHCb:2021awg, Belle:2018iff} and EDM experiments~\cite{ACME:2018yjb, nEDM:2020crw}.
We find several benchmark scenarios which can explain the BAU.
Finally, we discuss how we can test these benchmark scenarios at future collider experiments~\cite{Cepeda:2019klc, Bambade:2019fyw, Fujii:2015jha, CLICdp:2018cto}, various flavor experiments~\cite{Belle-II:2018jsg, LHCb:2012myk} and future gravitational wave observations~\cite{LISA:2017pwj, Seto:2001qf, Corbin:2005ny}.
In particular, the model can be tested by the di-photon decay of the Higgs boson~\cite{HiggsGamma_Early1,  HiggsGamma_Early2, Barroso:1999bf, Arhrib:2003vip, Djouadi:2005gj, Akeroyd:2007yh, Posch:2010hx} and the triple Higgs boson coupling due to the large deviation by non-decoupling effects which cause strongly first order phase transition~\cite{Kanemura:2002vm, Kanemura:2004ch, Kanemura:2004mg, Braathen_Kanemura}.
In the viable scenario with a relatively large wall velocity, enough amounts of gravitational waves~\cite{Grojean:2006bp, Caprini:2015zlo, Kakizaki:2015wua, Hashino:2016rvx, Espinosa:2010hh} can be predicted for the observations at future space-based interferometers.

This paper is organized as follows.
In section~\ref{sec:MODEL}, we define the two Higgs doublet model.
In section~\ref{sec:CONST}, we discuss theoretical and experimental constraints on the model.
In section~\ref{sec:EWPT}, the effective potential and some numerical results about electroweak phase transition are discussed in this model.
In section~\ref{sec:BAU}, the transport equations and the numerical results for the BAU are shown, and we give some predictions for future experiments in some benchmark points.
Some comments and phenomenological implications are discussed in section~\ref{sec:DISCUSS}, and conclusions are given in section~\ref{sec:CONCL}.

\section{The Model\label{sec:MODEL}}

In this paper, we discuss the THDM with two isospin doublets $\Phi_1$ and $\Phi_2$ with hypercharges $Y=1/2$.
Both the Higgs doublets can obtain the Vacuum Expectation Values (VEVs) which break the electroweak gauge symmetry. 
By a unitary transformation, we can choose the basis of the Higgs doublets so that only one of them has the VEV and the other does not (the Higgs basis)~\cite{Davidson:2005cw}. 
In the following, the Higgs basis is employed. 

In the Higgs basis, the elements of the Higgs doublets are represented by 
\begin{equation}
    \Phi_1 =
    \begin{pmatrix}
    G^+ \\ \frac{1}{\sqrt{2}}(v + h_1 + i G^0)
    \end{pmatrix}, ~~~~
    \Phi_2 =
    \begin{pmatrix}
    H^+ \\ \frac{1}{\sqrt{2}}(h_2 + i h_3)
    \end{pmatrix}. 
    \label{MODELhiggsbasis}
\end{equation} 
The scalar fields $G^0$ and $G^\pm$ are the Nambu-Goldstone modes. 
They are absorbed into the longitudinal modes of $W^\pm$ and $Z$ bosons, respectively. 
Other scalar fields are physical Higgs bosons. 
Therefore, this model has three additional Higgs bosons: two neutral ones and a pair of charged ones. 

The Higgs potential is given by 
\begin{align}
\mathcal{V} = &-{\mu_1}^2({\Phi_1}^\dagger \Phi_1) -{\mu_2}^2({\Phi_2}^\dagger \Phi_2) -\left( {\mu_3}^2({\Phi_1}^\dagger \Phi_2) + \mathrm{h.c.} \right) \notag \\[5pt]
&+\frac{1}{2}\lambda_1({\Phi_1}^\dagger \Phi_1)^2  +\frac{1}{2}\lambda_2({\Phi_2}^\dagger \Phi_2)^2 +\lambda_3({\Phi_1}^\dagger \Phi_1)({\Phi_2}^\dagger \Phi_2) +\lambda_4({\Phi_2}^\dagger \Phi_1)({\Phi_1}^\dagger \Phi_2) \notag \\[5pt]
&+ \left\{\left(\frac{1}{2}\lambda_5{\Phi_1}^\dagger \Phi_2 + \lambda_6{\Phi_1}^\dagger \Phi_1 + \lambda_7{\Phi_2}^\dagger \Phi_2 \right){\Phi_1}^\dagger \Phi_2 + \mathrm{h.c.}  \right\}.   
\end{align}
In general, $\mu_3^2$, $\lambda_5$, $\lambda_6$, and $\lambda_7$ are complex. 
One of them can be a real parameter by appropriately redefining the phase of the second Higgs doublet $\Phi_2$.  
Thus, three CP-violating phases are generally included in the Higgs potential. 

By substituting Eq.~(\ref{MODELhiggsbasis}) into the Higgs potential, 
the stationary conditions give
\begin{equation}
    \mu^2_1 = \frac{\lambda_1}{2}v^2, \quad \mu^2_3 = \frac{\lambda_6}{2}v^2. 
    \label{eq:station}
\end{equation}
The second condition in Eq. (\ref{eq:station}) means that the CP-violating phases of $\mu_3^2$ and $\lambda_6$ are the same. 
Only two of the three CP-violating phases are independent. 

The mass of $H^\pm$ is given by
\begin{equation}
m_{H^\pm}^2 = M^2 + \frac{ 1 }{ 2 } \lambda_3 v^2, 
\end{equation}
where $M^2 = - \mu_2^2$. The mass terms for the neutral scalar states are given by $\frac{ 1 }{ 2 } \sum_{k, \ell=1}^3 h_k h_\ell \mathcal{M}^2_{k\ell}$, 
where $\mathcal{M}^2$ is the three-by-three matrix defined by 
\begin{equation}
    \mathcal{M}^2 = v^2
    \begin{pmatrix}
    \lambda_1  & \mathrm{Re}[\lambda_6] & -\mathrm{Im}[\lambda_6] \\
    \mathrm{Re}[\lambda_6] & \frac{M^2}{v^2} + \frac{1}{2}(\lambda_3 + \lambda_4 + \mathrm{Re}[\lambda_5] ) & -\frac{1}{2} \mathrm{Im}[\lambda_5] \\
    -\mathrm{Im}[\lambda_6] & -\frac{1}{2}\mathrm{Im}[\lambda_5] & \frac{M^2}{v^2} + \frac{1}{2}(\lambda_3 + \lambda_4 - \mathrm{Re}[\lambda_5])
    \end{pmatrix}. 
    \label{eq: Neutral_mass_matrix}
\end{equation}
Since $\mathcal{M}^2$ is a symmetric matrix, it can be diagonalized by a real orthogonal matrix $R$. 
The mass eigenstates of the neutral scalar states are given by
\begin{equation}
H_k = \sum_{\ell=1}^3 R_{\ell k} h_\ell, \quad (k=1,2,3). 
\end{equation}

Non-diagonal elements of the mass matrix $\mathcal{M}^2$ induce the mixing among the neutral scalar states. 
The imaginary part of $\lambda_5$ can be zero by appropriately fixing the phase of $\Phi_2$. 
The mixing is then induced by only one scalar coupling $\lambda_6$. 

If $h_1$ is not a mass eigenstate but a linear combination of $H_k$ ($k=1,2,3$), 
the model predicts the tree-level induced deviation of the coupling constants of the $125~\mathrm{GeV}$ Higgs boson from its SM prediction. 
It is strongly constrained by the LHC results so far~\cite{ATLAS:2019nkf, CMS:2018uag}. 
To avoid them, we simply assume an alignment in the mass matrix, i.e. $\lambda_6$ is taken to be zero~\cite{Kanemura:2020ibp}. 
In the following, we call this simplification the Higgs alignment. 
The matrix $R$ is the identity matrix ($R_{\ell k} = \delta_{\ell k}$) in this case. 

In the Higgs alignment scenario, the masses of the neutral Higgs boson are given by
\begin{align}
    m_{H_1}^2 &= \lambda_1 v^2, \notag \\
    m_{H_2}^2 &= M^2 + \frac{\lambda_3 + \lambda_4 + \mathrm{Re}[\lambda_5]}{2} v^2, \notag \\
    m_{H_3}^2 &= M^2 + \frac{\lambda_3 + \lambda_4 - \mathrm{Re}[\lambda_5]}{2} v^2. 
\end{align}
We consider $H_1$ as the observed Higgs boson. 
Then, the coupling $\lambda_1$ is determined by the VEV $v=246~\mathrm{GeV}$ and the Higgs boson mass $m_{H_1} = 125~\mathrm{GeV}$. 
Remaining undetermined parameters in the Higgs potential are seven: $M^2$, $\lambda_2$, $m_{H^\pm}$, $m_{H_2}$, $m_{H_3}$, $|\lambda_7|$, and $\theta_7 \equiv \mathrm{arg}[\lambda_7]$.  
There is only one CP-violating parameter $\theta_7$ in the Higgs potential because we consider the Higgs alignment scenario. 

Next, the kinetic terms for the Higgs doublets are given by
\begin{equation}
    \mathcal{L}_{\mathrm{kin}} = |D_\mu \Phi_1|^2 + |D_\mu \Phi_2|^2, 
\end{equation}
where the covariant derivative $D_\mu$ defined as 
\begin{equation}
D_\mu \equiv \partial_\mu  - ig \frac{\sigma^a}{2} W_\mu^a - i\frac{1}{2} g^\prime B_\mu, 
\end{equation}
where $g$ and $g^\prime$ are the gauge coupling constants for $SU(2)_L$ and $U(1)_Y$, respectively. 
The $HVV$ interactions are given as follows:
\begin{equation}
    \mathcal{L}_{HVV} = \sum_{k=1}^3 R_{1k} \left(\frac{2m_W^2}{v} W^\dagger_\mu W^\mu + \frac{m_Z^2}{v} Z_\mu Z^\mu \right) H_k. 
    \label{MODELintHVV}
\end{equation}
In the Higgs alignment scenario, only $H_1$ has the $HVV$ interactions at the tree level because $R_{\ell k} = \delta_{\ell k}$. 
Their coupling constants ($H_1WW$ and $H_1ZZ$) coincide with those in the SM at the tree level. 

The Yukawa interaction is given by
\begin{equation}
    \mathcal{L}_{y} = - \sum_{k=1}^2 \sum_{i,j} \left( \overline{Q^\prime_{iL}} (y^k_u)_{ij}^\dagger \tilde{\Phi}_k u^\prime_{jR} + \overline{Q^\prime_{iL}} (y_d^k)_{ij} \Phi_k d^\prime_{jR} + \overline{L^\prime_{iL}} (y_e^k)_{ij} \Phi_k e^\prime_{jR} + \mathrm{h.c.} \right),
\end{equation}
where $\tilde{\Phi}_k$ ($k=1,2$) are defined as $\tilde{\Phi}_k=i\sigma_2 \Phi_k^\ast$. 
The fermion fields $Q_{iL}^\prime$ ($L_{iL}^\prime$) are the left-handed quark (lepton) doublets, where $i$ is the flavor index ($i=1,2,3$). 
The right-handed up-type quarks, down-type quarks and leptons are denoted by $u_{iR}^\prime$, $d_{iR}^\prime$ and $\ell_{iR}^\prime$, respectively. 

In general, two Yukawa matrices $y_f^1$ and $y_f^2$ ($f=u,d,e$) cannot be diagonalized simultaneously. 
However, flavor non-diagonal Yukawa couplings induce the dangerous FCNCs at the tree level \cite{Glashow:1976nt}, which are severely constrained by the flavor experiments so far. 
To avoid tree-level FCNCs, we assume the Yukawa alignment scenario~\cite{Pich:2009sp}, where the two Yukawa matrices are proportional to each other;  
\begin{equation}
    y_f^2 = \zeta_f y_f^1, \quad (f=u,d,e). 
\end{equation}
The coefficients $\zeta_f$ are complex. 

In the Yukawa alignment scenario, 
the interaction between the Higgs bosons and the SM fermions are given by
\begin{align}
    \mathcal{L}_{y} = &-\sum_{i,j} \left\{ \sum_{f=u,d,e} \sum_{k=1}^3 \overline{f_{iL}} \left(\frac{(M_f)_{ij}}{v} \kappa^k_f \right) f_{jR} H_k  \right. \notag \\
    &+ \left. \frac{\sqrt{2}}{v} \left\{ -\zeta_u \overline{u_{iR}}(M^\dagger_u V_{CKM})_{ij} d_{jL} + \zeta_d \overline{u_{iL}} (V_{CKM} M_d)_{ij} d_{jR} + \zeta_e \overline{\nu_{iL}} (M_e)_{ij} e_{jR} \right\} H^+ \right\} + \mathrm{h.c.} \notag \\
    &+ \cdots, 
    \label{MODELintHFF}
\end{align}
where the fermion fields without the prime (${}^\prime$) denote the mass eigenstates. 
$M_f$ ($f=u,d,e$) are the diagonal mass matrices defined as 
\begin{equation}
    M_u = \mathrm{diag}(m_u,m_c,m_t), \quad 
    M_d = \mathrm{diag}(m_d,m_s,m_b), \quad 
    M_e = \mathrm{diag}(m_e,m_\mu,m_\tau). 
\end{equation}
The coefficients $\kappa_f^k$ ($k=1,2$) are given by
\begin{equation}
\begin{cases}
    &\kappa_u^1 = 1, \quad \kappa_u^2 = \zeta_u^\ast, \quad \kappa_u^3 = -i\kappa_u^2, \\
    &\kappa_d^1 = 1, \quad \kappa_d^2 = \zeta_d, \quad \kappa_d^3 = i\kappa_d^2, \\
    &\kappa_e^1 = 1, \quad \kappa_e^2 = \zeta_e, \quad \kappa_e^3 = i\kappa_e^2. 
\end{cases}
\label{MODELdefkappa}
\end{equation}
All the $\kappa_f^1$ ($f = u, d, e$) is equal to unity because we consider the Higgs alignment scenario. 
The Yukawa interaction between $H_1$ and the SM fermions coincides with the SM one at the tree level. 

The interaction between the additional Higgs bosons and the SM fermions is described by six new real parameters $|\zeta_f|$ and $\theta_f \equiv \mathrm{arg}[\zeta_f]$ ($f=u,d,e$). 
The phases $\theta_f$ are the CP-violating phases. 
With a specific relation among the values of $\zeta_f$, 
the Yukawa interaction of the model coincides with that in the softly broken $Z_2$ symmetric two Higgs doublet model \cite{Glashow:1976nt, Barger:1989fj, Grossman:1994jb, Aoki:2009ha}, which is categorized into Type-I, Type-II, Type-X and Type-Y. 
In Table~\ref{MODELreration2HDM}, the values of $\zeta_f$ are shown in each type of Yukawa interaction \cite{Pich:2009sp}. 
\begin{table}[t]
\begin{tabular}{|c|c|c|c|}
\hline
 & $\quad\quad  \zeta_u\quad\quad$ & $\quad \quad \zeta_d\quad\quad $ & $\quad\quad \zeta_e\quad\quad$ \\ \hline
Type I & $\cot\beta$ & $\cot\beta$ & $\cot\beta$ \\ \hline
Type II & $\cot\beta$ & $-\tan\beta$ & $-\tan\beta$  \\ \hline
Type X & $\cot\beta$ & $\cot\beta$ & $-\tan\beta$ \\ \hline
Type Y & $\cot\beta$ & $-\tan\beta$ & $\cot\beta$ \\ \hline
\end{tabular}
\centering
\caption{The values of $\zeta_f ~ (f = u,d,e)$ in each type of Yukawa interaction}
\label{MODELreration2HDM}
\end{table}

\section{Constraints on the model \label{sec:CONST}}
In this section, we discuss parameter spaces of the model under theoretical and experimental constraints.
We consider perturbative unitarity and vacuum stability as the theoretical constraints.
We consider the experimental constraints from the collider, flavor and EDM data and also the electroweak precision tests.

\subsection{Theoretical constraints}
In this subsection, we consider the theoretical bounds in the model: perturbative unitarity, vacuum stability, and triviality.  
The constraint from perturbative unitarity in the THDMs has been investigated in various literature. 
In refs.~\cite{Kanemura:1993hm, Akeroyd:2000wc, Ginzburg:2005dt}, the perturbative unitarity bound has been studied in the THDMs with (softly broken) $Z_2$ symmetry~\cite{Glashow:1976nt}. 
The bound in the general THDMs (without $Z_2$ symmetry) has been investigated in ref.~\cite{Kanemura:2015ska}. 
We employ the formulae in ref.~\cite{Kanemura:2015ska} for the perturbative unitarity bound. 

Next, we consider the constraint from vacuum stability. 
The Higgs potential has to be bounded from below for the stability of the vacuum. 
This condition leads to the bounds on quartic scalar couplings in the THDMs that are given in refs.~\cite{Deshpande:1977rw, Klimenko:1984qx, Sher:1988mj, Sher:1988mj, Nie:1998yn, Ferreira:2004yd}. 
In the case that $\lambda_6 = \lambda_7 = 0$, i.e. the $Z_2$ symmetry is conserved in the quartic terms in the Higgs potential, 
the condition yields
\begin{equation}
 \lambda_1 \ge 0, \quad \lambda_2 \ge 0, \quad \sqrt{\lambda_1 \lambda_2} + \lambda_3 \ge 0, \quad 
  \sqrt{\lambda_1 \lambda_2} + \lambda_3 + \lambda_4 \pm \mathrm{Re}[\lambda_5] \ge 0.
  \label{eq:bounded1} 
 \end{equation}
Eq. (\ref{eq:bounded1}) is not only the necessary condition but also the sufficient condition~\cite{Klimenko:1984qx}. 
In the general THDM with the Higgs alignment scenario, in addition, we employ the following necessary conditions according to discussion in ref.~\cite{Ferreira:2004yd}; 
\begin{equation}
\begin{array}{l}
  \displaystyle{\Bigl| \mathrm{Re}[\lambda_7] \Bigr| \le \frac{1}{4}(\lambda_1 +\lambda_2) + \frac{1}{2}(\lambda_3 +\lambda_4 + \mathrm{Re}[\lambda_5])}, \\[12pt]
  \displaystyle{ \Bigl|\mathrm{Im}[\lambda_7] \Bigr| \le \frac{1}{4}(\lambda_1 +\lambda_2) + \frac{1}{2}(\lambda_3 +\lambda_4 - \mathrm{Re}[\lambda_5])}. \\
  \end{array}
  \end{equation}

Finally, the scalar coupling constants in the Higgs potential are also constrained as a function of the cut-off scale $\Lambda$ by the renormalization group equation analysis~\cite{Lindner:1985uk}, where we impose condition that the running coupling constants do not blow up nor fall down below $\Lambda$.
Imposing that the running couplings are smaller than a critical value (usually being set to be $4\pi$) up to $\Lambda$, 
the upper and lower limits of the magnitudes of the scalar coupling constants are obtained.  
In the THDMs, they have been investigated in refs.~\cite{Kanemura:1999xf, Flores:1982pr, Kominis:1993zc, Ferreira:2009jb, Cline:2011mm, Dorsch:2016nrg}. 
In general, in extended scalar models positive additional terms are added to beta functions of scalar coupling constants.
Therefore the scalar coupling constants tend to blow up.
Consequently, $\Lambda$ can appear at relatively lower scales as the Landau pole.
In order to keep the Landau pole to be above TeV scales, the scalar coupling constants at the electroweak scale are constrained. 
In refs.~\cite{Cline:2011mm, Dorsch:2016nrg}, 
such a bound on the coupling constants is discussed in the context of EWBG.
We here do not explicitly study these renormalization group analyses as it is out of scope of this paper.
Instead, we give comments on this issue in section \ref{sec:DISCUSS}.

\subsection{Constraints from collider and flavor experiments}
\subsubsection{Collider experiments}
In this subsection, we consider constraints from collider data.
First, we discuss the direct search experiments for charged Higgs bosons at LEP and LHC.
From the result at the LEP experiment~\cite{ALEPH:2013htx}, a lower bound of the mass is given by $m_{H^\pm} \gtrsim 80$ GeV almost independent of $\zeta_f~(f = u,d,e)$.
When the mass region is $80 ~\mathrm{GeV} \lesssim m_{H^\pm} \lesssim 170 ~\mathrm{GeV}$, the charged Higgs bosons are produced in the top quark decay process $t \to H^\pm b$.
However, the upper bound of $\mathcal{B}(t \to H^\pm b) \times \mathcal{B}(H^\pm \to \tau^\pm \nu)$ in this mass region is severely constrained from ATLAS~\cite{ATLAS:2018gfm} and CMS data~\cite{CMS:2019bfg}, and the branching ratio needs to satisfy $\mathcal{B}(t \to H^\pm b) \lesssim O(10^{-3})$ when $\mathcal{B}(H^\pm \to \tau^\pm \nu) = 1$~\cite{Kanemura:2020ibp}.
Therefore, we only consider the mass region $m_{H^\pm}>m_t$ in the following discussions.
In this case, a leading production process of the charged Higgs bosons is $gb \to t H^\pm$.
They mainly decay into $tb$ or $\tau \nu$, and can also decay into an off-shell $W$ boson and a neutral Higgs boson if it is kinematically allowed~\cite{Kanemura:2020ibp, Kanemura:2021dez}.
$\zeta_f$ are constrained from $H^\pm \to tb$~\cite{ATLAS:2021upq} and $H^\pm \to \tau^\pm \nu$~\cite{CMS:2019bfg} searches.
In our analysis, we replace $1 /\tan\beta$ to $|\zeta_u|$ in the production cross section $\sigma(gb\to tH^\pm)$ in the case of Type-I THDM as long as $|\zeta_d|$ is not too large.\footnote{In the production of $H^\pm$, we neglect the effects of the CP phases. }
We have referred to the value of $\sigma(gb\to tH^\pm)$ from figure 9 in ref.~\cite{Aiko:2020ksl}.

Second, we discuss oblique parameters such as $S, T, U$~\cite{oblique_parameter}, especially the $T$ parameter.
$\lambda_4 - \mathrm{Re}[\lambda_5]$ and $\mathrm{Im}[\lambda_7]$ terms in the potential violate the custordial symmetry~\cite{Sikivie:1980hm, Haber:1992py, Pomarol:1993mu, Gerard:2007kn, Haber:2010bw, Grzadkowski:2010dj, Aiko:2020atr}.
Consequently, the existence of these terms causes a deviation in the $T$ parameter from the SM value, and this is constrained from the electroweak fitting results~\cite{Baak:2012kk}.
To avoid this, we assume $m_{H^\pm} = m_{H_3}$ in our analysis, since $\lambda_4 - \mathrm{Re}[\lambda_5]$ is proportional to $m_{H_3}^2 - m_{H^\pm}^2$.
Under this assumption, the effects of the heavy Higgs bosons involving $\mathrm{Im}[\lambda_7]$ do not contribute to the $T$ parameter at one loop level~\cite{Pomarol:1993mu, Haber:2010bw}.

Third, we discuss the direct searches for neutral Higgs bosons $H_2$ and $H_3$ at LHC.
There are three single production processes of the neutral Higgs bosons such as $gg \to H_{2,3}$ (gluon fusion), $g g \to H_{2,3}b\overline{b}$ (bottom associated) and $g g \to H_{2,3} t \overline{t}$ (top associated).
These production cross sections are given by~\cite{Bian:2017jpt, Kanemura:2020ibp}
\begin{align}
    &\sigma (gg \to H_{2,3}) = \sigma (gg \to H_1)_{\mathrm{SM}} \times \frac{\Gamma (H_{2,3} \to gg)}{\Gamma (H_{2,3} \to gg)_{\kappa^{2,3}_u=\kappa^{2,3}_d=1}},\notag \\
    &\sigma (gg \to H_{2,3}b\overline{b}) = |\zeta_d|^2 \sigma(gg \to H_1 b\overline{b})_{\mathrm{SM}}, \notag \\
    &\sigma (gg \to H_{2,3}t\overline{t}) = |\zeta_u|^2 \sigma(gg \to H_1 t\overline{t})_{\mathrm{SM}},
    \label{eq:neutprod}
\end{align}
where $\sigma(gg\to H_1)_{\mathrm{SM}}$ and $\sigma(gg\to H_1 b\overline{b})_{\mathrm{SM}}$ ($\sigma(gg\to H_1 t \overline{t})_{\mathrm{SM}}$) are the production cross section of the SM Higgs boson for gluon fusion and bottom (top) associated, respectively.
We have referred to the value of the cross section for the gluon fusion process (NNLO in QCD) and bottom and top associated processes (NLO in QCD) from ref.~\cite{LHC_HIGGS_WG}.
We consider decays of the neutral Higgs bosons into a fermion pair ($H_{2,3} \to t\overline{t}, b\overline{b}, \tau \overline{\tau}$), and loop-induced decays into a gluon pair ($H_{2,3} \to gg$).
$\Gamma(H_{2,3} \to gg)$ in Eq.~(\ref{eq:neutprod}) are the partial decay widths of $H_{2,3} \to gg$.
When one of the heavy Higgs bosons is heavier than the others, it can also decay into an off-shell gauge boson and another heavy Higgs boson.
In our benchmark points we will discuss below, the other decay modes are negligibly small~\cite{Kanemura:2020ibp, Kanemura:2021dez}.
When $m_{H_{2,3}}<2m_t$, $\zeta_f~(f = u,d,e)$ are constrained from the latest results of $H_{2,3} \to \tau \overline{\tau}$ searches by ATLAS~\cite{ATLAS:2020zms}.
On the other hand, when $m_{H_{2,3}}>2m_t$, the decay into a top quark pair is kinematically allowed, and
$\zeta_f $ are constrained from the current data of $H_{2,3} \to t\overline{t}$ searches by ATLAS~\cite{ATLAS:2018rvc} and CMS~\cite{CMS:2019pzc}.

Recently, in the aligned THDM without CP phases, constraints on electroweak pair productions of the heavy Higgs bosons~\cite{Kanemura:2001hz, Cao:2003tr, Belyaev:2006rf} were discussed in ref.~\cite{Kanemura:2021dez}.
Events including multi tau leptons in the final state have been searched at the LHC~\cite{ATLAS:2021yyr}.
There are six types of the pair production process in the model :
\begin{equation}
    pp \to H_2 H_3,~~ pp \to H_2 H^\pm, ~~ pp \to H_3 H^\pm,~~ pp \to H^+ H^-.
\end{equation}
According to ref.~\cite{Kanemura:2021dez}, the constraints on $|\zeta_d|$ and $|\zeta_e|$ from the decay processes $H_2 H_3  \to 4 \ell$ including multi tau leptons in final state become the most severe at $|\zeta_u| = 0$ due to the suppression of $H_{2,3} \to c \overline{c}$ and $H^\pm \to tb$.
As shown in figure 9 in ref.~\cite{Kanemura:2021dez}, when $m_{H_2}=m_{H_3}=m_{H^\pm} \equiv m_{\Phi} = 280$ GeV, the region $|\zeta_d| \lesssim 0.1 \times |\zeta_e|$ is excluded from the multi lepton search.
When the charged Higgs bosons are heavier than the neutral Higgs boson, a decay channel into a tau pair via the decay into the off-shell $W$ boson and a neutral Higgs boson is open.
As a result, the excluded region in the $|\zeta_d|$-$|\zeta_e|$ plane becomes large compared to the case of the light charged Higgs bosons.
We note that there are no constraint on the region $|\zeta_d| \gtrsim | \zeta_e|$ from the multi lepton search.

\subsubsection{Flavor experiments}
Fourth, we discuss flavor experiments and their impacts on the parameter space in the aligned THDM.
In addition to the SM contributions, diagrams involving the additional Higgs boson exchanges contribute to $B \to X_s \gamma,~ B_{d,s} \to \mu \mu$ and leptonic tau decays.
The current experimental value of the branching ratio for $B \to X_s \gamma$ with the photon energy cut $E_\gamma > 1.6$ GeV is given by~\cite{HFLAV:2019otj, Haller:2018nnx}
\begin{equation}
\mathcal{B}(B \to X_s \gamma)_{\mathrm{exp}} = (3.32 \pm 0.15) \times 10^{-4},
\label{eq:BSGEXP}
\end{equation}
which is the combined result from BABAR~\cite{BSG_BaBar}, Belle~\cite{BSG_Belle} and CLEO~\cite{CLEO:2001gsa}.
In the aligned THDM, $\zeta_u$, $\zeta_d$ and $m_{H^\pm}$ are constrained from Eq.~(\ref{eq:BSGEXP}).
In the SM, the branching ratio is given by $\mathcal{B}(B \to X_s \gamma)_{\mathrm{SM}} = (3.36 \pm 0.23) \times 10^{-4}$ at NNLO in QCD~\cite{Czakon:2015exa} with the same photon energy cut.
According to ref.~\cite{Modak_Senaha}, we define 
\begin{equation}
\mathcal{B}(B \to X_s \gamma)_{\mathrm{th}} = R_{\mathrm{th}} \cdot \mathcal{B}(B \to X_s \gamma)_{\mathrm{SM}},
\end{equation}
where,
\begin{equation}
R_{\mathrm{th}} = \frac{\mathcal{B}(B \to X_s \gamma)_{\mathrm{THDM}}}{\mathcal{B}(B \to X_s \gamma)_{\zeta_u=\zeta_d=0 }},
\end{equation}
as a prediction in the aligned THDM.
We have used the formulae given in~\cite{Borzumati:1998tg} at NLO in QCD for the calculation of $\mathcal{B}(B \to X_s \gamma)_{\mathrm{THDM}}$ and $\mathcal{B}(B \to X_s \gamma)_{\zeta_u=\zeta_d=0}$.

Observed values of the branching ratios of $B_{d} \to \mu \mu$ and $B_{s} \to \mu \mu$ can be referred in refs.~\cite{HFLAV:2019otj, ATLAS:2018cur, CMS:2019bbr, LHCb:2021awg}:
\begin{align}
&\mathcal{B} (B_d \to \mu \mu) < 2.1 \times 10^{-10} ~(95 \%~ \mathrm{C. L.}), \notag \\
&\mathcal{B}(B_s \to \mu \mu) = (3.1 \pm 0.6) \times 10^{-9}.
\label{eq:BMUEXP}
\end{align}
In the aligned THDM, $\zeta_f$ ($f = u,d,e$), $m_{H_{2,3}}$ and $m_{H^\pm}$ are constrained from Eq.~(\ref{eq:BMUEXP}).
The decay rates of $B_{d,s} \to \mu \mu$ in the aligned THDM were calculated by ref.~\cite{Li:2014fea}, and we use the value of the SM contribution of the Wilson coefficient $C_{10}$ of the operator $(\overline{q} \gamma_\mu P_L b)(\overline{l} \gamma^\mu \gamma_5 l)$ as 
\begin{equation}
C_{10}^{\mathrm{SM}} = -0.938 \times \left( \frac{m_t}{173.1~\mathrm{GeV}} \right)^{1.53} \times \left( \frac{\alpha_s(m_Z)}{0.1184} \right)^{-0.09},
\end{equation}
which is evaluated at NNLO in QCD~\cite{Bobeth:2013uxa}.
Regarding $B \to X_s \gamma$ and $B_{d,s} \to \mu \mu$, we require these theoretical values to be within the 2$\sigma$ deviations from the experimental data.

According to ref.~\cite{Kanemura:2021dez}, $\zeta_e$ and $m_{H^\pm}$ are constrained from the leptonic tau decay in the model.
One can see that $|\zeta_e| \gtrsim 70$ (90) are excluded with $m_{H^\pm} = 180$ (230) GeV from figure~6 in ref.~\cite{Kanemura:2021dez}.

\begin{figure}[t]
    \centering
    \includegraphics[width=1.0\linewidth]{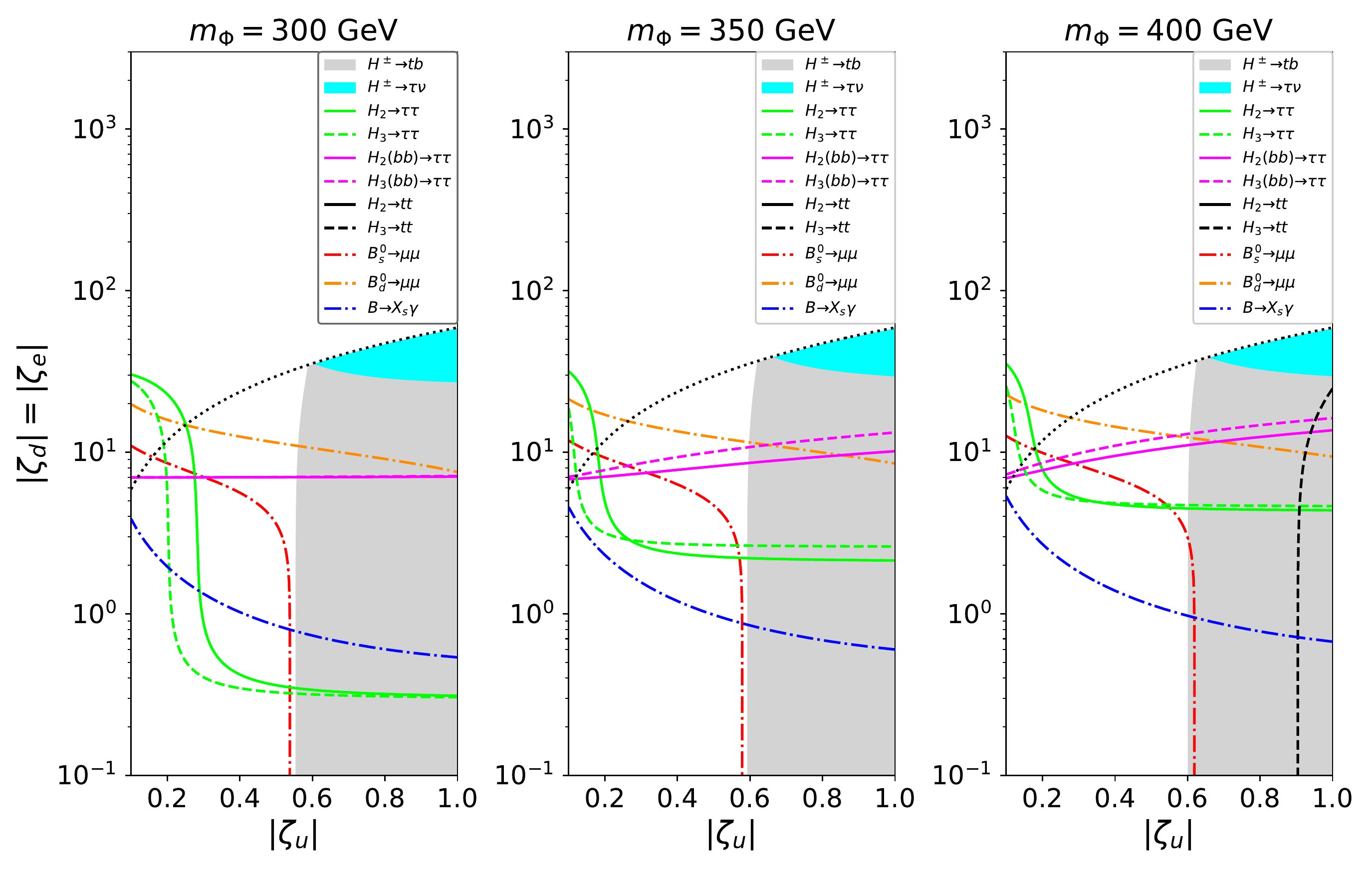}
    \caption{The constraints on $|\zeta_f|~(f = u,d,e)$ from direct searches and flavor experiments. Input parameters are set by $M = 30~\mathrm{GeV}, ~\lambda_7 = 0.8, ~\theta_7 = -0.9,~ \theta_u = \theta_d= -2.7,~\theta_e = \theta_u + 0.04$, and the heavy Higgs masses $m_{\Phi}$ are degenerated: $m_{\Phi} = 300$ GeV (left), $m_{\Phi} = 350$ GeV (middle), $m_{\Phi} = 400$ GeV (right). Black dotted lines satisfy $|\zeta_d| = (m_t/m_b) |\zeta_u|$, and the gray (cyan) regions below these lines are excluded from $H^\pm \to t b$, ($H^\pm \to \tau \nu$). The green solid (dashed) lines and the magenta solid (dashed) lines are the upper bounds from $H_2 \to \tau \overline{\tau}$, ($H_3 \to \tau \overline{\tau}$) and $H_2 (b\overline{b}) \to \tau \overline{\tau}$, ($H_3 (b\overline{b}) \to \tau \overline{\tau}$), respectively. In the right panel, the constraint from $H_3 \to t \overline{t}$ can be seen as the black dashed line. The regions above the red (orange) and blue dashdot lines are excluded by $B_{s} (B_d) \to \mu\mu$ and $B \to X_s \gamma$, respectively.}
    \label{fig:ConstDirect}
\end{figure}
The constraints on $|\zeta_f|$ from direct searches and flavor experiments are shown in figure~\ref{fig:ConstDirect}.
The SM input parameters we have used are shown in Table~\ref{table:SMinputs}.
The heavy Higgs masses $m_{\Phi}$ are degenerated,  and the left, middle and right panels are in the cases of $m_{\Phi} = 300,  350$ and $400$ GeV, respectively.
We consider the case of $|\zeta_d| = |\zeta_e|$, and other relevant input parameters in the model are set by $M = 30~\mathrm{GeV}, ~\lambda_7 = 0.8, ~\theta_7 = -0.9,~ \theta_u = \theta_d = -2.7, ~\theta_e = \theta_u +0.04$.
In figure~\ref{fig:ConstDirect}, black dotted lines satisfy $|\zeta_d| = (m_t/m_b) |\zeta_u|$ which are the reliable bounds about the calculation of $\sigma(gb\to tH^\pm)$, and the gray (cyan) regions below these lines are excluded from $H^\pm \to t b$, ($H^\pm \to \tau \nu$).
The green solid (dashed) lines and the magenta solid (dashed) lines are the upper bounds from $H_2 \to \tau \overline{\tau}$, ($H_3 \to \tau \overline{\tau}$) and $H_2 (b\overline{b}) \to \tau \overline{\tau}$, ($H_3 (b\overline{b}) \to \tau \overline{\tau}$), respectively.
In the right panel of figure~\ref{fig:ConstDirect} which is the case of $m_{\Phi} = 400$ GeV, the decay channel to a top quark pair is kinematically allowed.
Thus, the constraint from $H_3 \to t \overline{t}$ can be seen as the black dashed line.
The regions above the red (orange) and blue dashdot lines are excluded from $B_{s} (B_d) \to \mu\mu$ and $B \to X_s \gamma$, respectively.

In these mass regions, we found that $|\zeta_u| \gtrsim 0.6$ are excluded from $H^\pm \to t b$ and $B_s \to \mu \mu$ almost independent of the heavy Higgs boson masses. 
In the left panel of figure~\ref{fig:ConstDirect}, $|\zeta_d|=|\zeta_e| \gtrsim 4$ is excluded from $B \to X_s \gamma$ with $|\zeta_u| \lesssim 0.2$, while $|\zeta_d|=|\zeta_e| \gtrsim 0.3$  is excluded from $gg \to H_{2,3} \to \tau \overline{\tau}$ with $|\zeta_u| \gtrsim 0.2$, because the production cross section $\sigma(gg \to H_{2,3})$ is large.
In the middle and right panels, only $B \to X_s \gamma$ sets the upper bound on $|\zeta_d|=|\zeta_e|$, and then we find that this behavior is almost irrelevant to the mass.

\begin{table}[t]
\begin{center}
\begin{tabular}{| l l l l |} \hline
$m_u = 1.29 \times 10^{-3}$, \quad & $m_c = 0.619$, \quad & $m_t = 171.7$, \quad & $m_W^{} = 80.379$,  \\
$m_d = 2.93 \times 10^{-3}$, \quad & $m_s = 0.055$, \quad & $m_b = 2.89$, \quad & $m_Z^{} = 91.1876$,\\
$m_e = 4.87 \times 10^{-4}$, \quad & $m_\mu = 0.103$, \quad & $m_\tau = 1.746$ \quad & (in GeV)~\cite{Xing:2007fb, Bijnens:2011gd, ParticleDataGroup:2020ssz}.  \\ \hline
$\lambda = 0.22453$, & $A = 0.836$,  & \qquad $\overline{\rho} = 0.122$, & \qquad\quad $\overline{\eta} = 0.355$, \\ \hline
$\alpha = 1/127.955$, & $\alpha_S = 0.1179$ & ~\cite{ParticleDataGroup:2020ssz}.&  \\ \hline
\end{tabular}
\end{center}
\begin{center}
\begin{tabular}{cccccccc} \hline \hline
~$m_{B_s}$ [GeV] & ~$m_{B_d}$ [GeV] & ~$f_{B_s} $ [GeV] & ~$f_{B_d}$ [GeV] & ~$\Gamma_{B_{sL}}^{-1}$ [ps] & ~$\Gamma_{B_{sH}}^{-1}$ [ps] & ~$\Gamma_{B_{dL}}^{-1}$ [ps] & ~$\mathrm{Br}(B \to X_c  l^- \nu)$ \\ \hline
$5.367$~\cite{ParticleDataGroup:2020ssz} & $5.280$~\cite{ParticleDataGroup:2020ssz} & $0.2284$~\cite{FlavourLatticeAveragingGroup:2019iem} & $0.1920$~\cite{FlavourLatticeAveragingGroup:2019iem} & $1.423$~\cite{ParticleDataGroup:2020ssz} &  $1.619$~\cite{ParticleDataGroup:2020ssz}  & $1.519$~\cite{ParticleDataGroup:2020ssz} &  $0.1065$~\cite{HFLAV:2019otj} \\ \hline \hline
\end{tabular}
\caption{The input parameters of the SM. The masses of the SM fermions and gauge bosons shown in the above table are the values at the $Z$ boson scale, and they are given in GeV. The symbols $\lambda$, $A$, $\overline{\rho}$ and $\overline{\eta}$ are the Wolfenstein parameters~\cite{Wolfenstein:1983yz}. The coupling constants $\alpha$ and $\alpha_S$ are the fine structure constant of QED and QCD, respectively. The below table shows the input parameters for the analyses of the flavor constraints. }
\label{table:SMinputs}
\end{center}
\end{table}

\subsection{EDM experiments}
We discuss constraints from EDM experiments in this subsection.
CP violation of the model is highly constrained from the electron and neutron EDM experiments.
The EDM $d_f$ of the fermion $f$ is defined as the coefficient of the effective operator, and this is written by 
\begin{equation}
    \mathcal{L}_{\mathrm{EDM}} = -\frac{d_f}{2}\overline{f}\sigma^{\mu \nu}(i \gamma_5) f F_{\mu\nu},
\end{equation}
where $F_{\mu \nu}$ is the field strength of electromagnetic fields and $\sigma^{\mu \nu} = \frac{i}{2}[\gamma^\mu,\gamma^\nu]$.

The current bound of the electron EDM (eEDM) from ACME~\cite{ACME:2018yjb} is given by $|d_e+kC_s|<1.1 \times 10^{-29}~e$  cm at 90\% C.L., where $C_s$ is the coefficient of the dimension six operator $\overline{e}i\gamma_5 e \overline{N} N$, which describes the interaction between electrons and nucleons, and the constant $k$ is about $\mathcal{O}(10^{-15})~ \mathrm{GeV}^2~e$ cm.
In our benchmarks we will discuss below, the contribution to the eEDM from $k C_s$ is about two orders of magnitude smaller than the current bound.
Therefore, we neglect this contribution, and we set the bound from the eEDM as $|d_e| < 1.0 \times 10^{-29}~e$ cm in the following discussions.

\begin{figure}[t]
    \centering
    \includegraphics[width=0.35\linewidth]{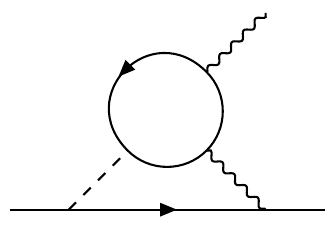}
    \includegraphics[width=0.35\linewidth]{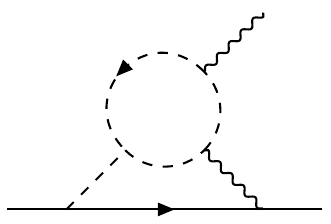}
    \caption{The Barr--Zee type diagrams including the fermion loop (left) and the scalar loop (right).} 
    \label{fig:BZdiagram}
\end{figure}
Two loop Barr--Zee type diagrams in figure~\ref{fig:BZdiagram} are leading contributions to the eEDM in the aligned THDM.
The left diagram in figure~\ref{fig:BZdiagram} has a fermion loop, and the right one has a scalar loop.  
Since there are multiple CP-violating phases in the model, each diagram depends on different CP-violating phases.
In the aligned THDM, the Barr--Zee type diagrams including the gauge boson loop do not exist, because of the condition of Higgs alignment~\cite{Kanemura:2020ibp}.
We can further categorize the Barr--Zee type diagrams depending on the scalar boson which couples to the external fermion line being either neutral or charged.

In calculation of the fermion loop contributions, we only consider the top quark loop diagrams because of the hierarchy in the Yukawa coupling constants ($y_t \gg y_b \gg y_\tau$).
Therefore, when $|\zeta_u|,|\zeta_d|$ and $|\zeta_e|$ are in the same order, the contributions from the fermion loop diagrams are approximately proportional to $|\zeta_u||\zeta_e| \sin( \theta_u -\theta_e)$.
On the other hand, the contributions from the heavy scalar loop diagrams are approximately proportional to $|\lambda_7||\zeta_e| \sin(\theta_7 - \theta_e)$.
CP-violating phases required to create the BAU can be $\mathcal{O} (1)$ under the eEDM constraint by the destructive interference between these independent diagrams.

We next discuss the neutron EDM (nEDM).
The most stringent constraint on the nEDM is $|d_n| < 1.8 \times 10^{-26}~ e$ cm at 90 \% C.L. by the NEDM collaboration~\cite{nEDM:2020crw}.
By using the QCD sum rule, $d_n$ is given by~\cite{Pospelov:2000bw, Hisano:2012sc, Fuyuto:2013gla, Abe:2013qla}
\begin{equation}
    d_n = 0.79 d_d - 0.20 d_u + e (0.59 d_d^C + 0.30 d_u^C) / g_3,
    \label{CONSTqcdsum}
\end{equation}
where $g_3$ is the coupling constant of the strong interaction, and $d_q^C~(q=u,d)$ is the chromo EDM.
In the case of the nEDM, contributions from the Weinberg operator $d_n(C_W)$~\cite{Weinberg:1989dx, Dicus:1989va} and the four fermi interaction~\cite{Khatsimovsky:1987fr} must be considered.
In the parameter regions which we discuss later, the order of magnitude of $d_n(C_W)$ is comparable to $d_n$,\footnote{Since the sign of $d_n(C_W)$ has theoretical uncertainties~\cite{Demir:2002gg, Jung:2013hka}, we consider both $d_n + d_n(C_W)$ and $d_n - d_n(C_W)$. } while the contributions from the four fermi interaction are negligibly small~\cite{Jung:2013hka}.
Therefore, in the calculation of the nEDM, we only consider $d_n$ and the contributions from the Weinberg operator $d_n(C_W)$.
According to the formulae shown in refs.~\cite{Boyd:1990bx, Chang:1990dja, Demir:2002gg, Abe:2013qla, Jung:2013hka, Kanemura:2020ibp}, when $m_{H_2}=m_{H_3}$, the nEDM is approximately proportional to $|\zeta_u||\zeta_d|\sin(\theta_u - \theta_d)$.

\section{Electroweak phase transition\label{sec:EWPT}}

In this section, the electroweak phase transition is discussed in the aligned THDM. 
First, the effective potentials are discussed at zero and finite temperatures. 
Second, we discuss profiles for the vacuum bubble generated at the electroweak phase transition. 
The results of numerical evaluations are shown.
We also show some formulae to find the bubble profiles.

\subsection{Effective potential of the model\label{subsec:EFFPOT}}

In this subsection, we show the effective potential of the configuration of the neutral elements $\varphi_1$, $\varphi_2$, and $\varphi_3$, 
which are defined as 
\begin{equation}
\label{eq: Classical_fields}
\left< \Phi_1 \right> = \frac{ 1 }{ \sqrt{2} } 
\begin{pmatrix}
0 \\
\varphi_1 \\
\end{pmatrix}, \quad 
\left< \Phi_2 \right> = \frac{ 1 }{ \sqrt{2} }
\begin{pmatrix}
0 \\
\varphi_2 + i \varphi_3 \\
\end{pmatrix}. 
\end{equation}
The imaginary part of $\left< \Phi_1 \right>$ can be set to zero by the gauge fixing.  
By substituting Eq.~(\ref{eq: Classical_fields}) into the Higgs potential, 
we obtain the tree-level effective potential. 

At the one-loop level, the effective potential at zero temperature is given by 
\begin{equation}
\label{eq: one-loop_Veff}
    V_{T=0}(\varphi_1, \varphi_2, \varphi_3) = V_0 + V_1 + V_{CT},
\end{equation}
where $V_0$ is the tree-level effective potential. 
The Coleman-Weinberg potential~\cite{Coleman:1973jx} with the Landau gauge is denoted by $V_1$; 
\begin{equation}
V_1 = \sum_k (-1)^{s_k} \frac{n_k}{64 \pi^2}\tilde{m}^4_k \left[ \log \frac{\tilde{m}_k^2}{Q^2} - \frac{3}{2} \right],
\end{equation}
where $Q$ is the renormalization scale. 
The index $k$ represents particles in one loop diagrams; the top quark $t$, weak bosons $W^\pm$ and $Z$, the photon $\gamma$, and the scalar bosons $G^\pm$, $G^0$, $H^\pm$, $H_1$, $H_2$, and $H_3$. 
We do not consider other particles because their effects are negligibly small. 
The degree of freedom of the particle $k$ is denoted by $n_k$. 
The fermion-loop diagram has the opposite sign of the boson-loop diagram. 
This difference is described by the factor $(-1)^{s_k}$, which is defined as $1$ ($-1$) for bosons (fermions). 
The field-dependent mass of the particle $k$ is given by $\tilde{m}_k$. 
The formulae of $\tilde{m}_k$ for each particle are shown for each particle in Appendix~\ref{sec:APP}. 

The counterterms are denoted by $V_{CT}$ in Eq.~(\ref{eq: one-loop_Veff}). 
To fix them, we employ the following nine renormalization conditions. 
\begin{align}
\label{eq: renormalization_cond_1}
   &\left. \frac{\partial V_{T=0} }{\partial \varphi_i} \right|_{(\varphi_1, \varphi_2, \varphi_3)=(v,0,0)} = 0, \quad (i = 1,2,3),\\
\label{eq: renormalization_cond_2}
   &\left. \frac{\partial^2 V_{T=0} }{\partial \varphi_i \partial \varphi_j} \right|_{(\varphi_1, \varphi_2, \varphi_3)=(v,0,0)} = \mathcal{M}_{ij}^2, 
 \quad (i,j=1,2,3;\ i\geq j), 
\end{align}
where $\mathcal{M}^2$ is the mass matrix for the neutral Higgs bosons in Eq.~(\ref{eq: Neutral_mass_matrix}). 
In evaluating the second derivative in Eq.~(\ref{eq: renormalization_cond_2}), 
Infrared Red (IR) divergences appear caused by the NG bosons. 
We set the IR cut-off scale to be $1~\mathrm{GeV}$ to avoid this difficulty~\cite{Baum:2020vfl}.   

Counterterms can be determined by conditions in Eqs.~(\ref{eq: renormalization_cond_1}) and~(\ref{eq: renormalization_cond_2}) except for those of $\mu_2^2, \lambda_2, \lambda_7$. 
The remaining three counterterms are fixed by the $\overline{\mathrm{MS}}$ scheme. 
The formulae for each counterterm are shown in Appendix~\ref{sec:APP}. 

By using $V_{T=0}$, the triple Higgs boson coupling $\lambda_{hhh}$ is evaluated as~\cite{Kanemura:2002vm, Kanemura:2004mg}
\begin{equation}
    \lambda_{hhh} = \left. \frac{\partial^3 V_{T=0}}{\partial \varphi_1^3} \right|_{(\varphi_1, \varphi_2, \varphi_3)=(v,0,0)},
\end{equation}
It is known that $\lambda_{hhh}$ is enhanced by the non-decoupling effect of the additional Higgs bosons~\cite{Kanemura:2002vm, Kanemura:2004ch, Kanemura:2004mg, Braathen_Kanemura}. 
The deviation in $\lambda_{hhh}$ from the SM prediction $\Delta R\equiv \lambda_{hhh}/ \lambda_{hhh}^\mathrm{SM} - 1$ is given as follows at one-loop level; 
\begin{equation}
    \Delta R = \frac{1}{12 \pi^2 v^2 m_{H_1}^2} \left\{ 2\frac{(m_{H^\pm}^2 - M^2)^3}{m_{H^\pm}^2} + \frac{(m_{H_2}^2 - M^2)^3}{m_{H_2}^2} + \frac{(m_{H_3}^2 - M^2)^3}{m_{H_3}^2} \right\}. 
    \label{eq:TRIPLECOUPLING}
\end{equation}

At finite temperature, the effective potential obtains thermal corrections.  
It is given by
\begin{equation}
\label{eq: Veff_Tneq0}
    V (\varphi_1, \varphi_2, \varphi_3; T) = V_{T=0} + V_T. 
\end{equation}
The term $V_T$ denotes the thermal correction evaluated by 
\begin{equation}
\label{eq: thermal_correction}
    V_T = \sum_{s_k} (-1)^{s_k} \frac{n_k}{2\pi^2 \beta^4} \int_0^\infty dx~x^2 \log \left( 1 + (-1)^{s_k+1} \exp \left( -\sqrt{x^2+\beta^2 \tilde{m}_k^2} \right)  \right),
\end{equation}
where $\beta = 1/T$ is the inverse temperature~\cite{Dolan:1973qd}.  
We employ the Parwani scheme~\cite{Parwani:1991gq} for thermal resummation, where the field-dependent mass also obtains the thermal corrections. 
Therefore, the field-dependent mass $\tilde{m}_k$ in Eq.~(\ref{eq: Veff_Tneq0}) includes the thermal correction. 
The formulae for the thermal correction of each $\tilde{m}_k$ are also shown in Appendix~\ref{sec:APP}. 

\subsection{Bubble profiles}
\label{subsec:BPROF}

In this subsection, bubble profiles of the electroweak phase transition are discussed. 
The behavior of the phase transition is investigated by using the effective potential at finite temperatures. 

The probability of tunneling at the temperature $T$ per unit time per unit volume is given by~\cite{Coleman:1977py, Callan:1977pt, Linde:1980tt} 
\begin{equation}
\Gamma = A(T) \exp\left(- \frac{ S_3 }{ T } \right), 
\end{equation}
where pre-factor $A(T)$ is roughly evaluated as $A(T) \sim T^4$ by the dimensional analysis. 
The probability is mainly determined by a three-dimensional Euclidian action $S_3$. 

The Euclidian action is calculated with O(3) symmetric solutions for the configurations $\varphi_1$, $\varphi_2$, and $\varphi_3$ determined by differential equations~\cite{Linde:1980tt}
\begin{equation}
\label{eq: bubble_equation}
\frac{ \mathrm{d}^2 \varphi_i }{ \mathrm{d} r^2 } + \frac{ 2 }{ r } \frac{ \mathrm{d} \varphi_i }{ \mathrm{d} r } = \frac{ \partial V }{ \partial \varphi_i }, \quad ( i=1,2,3), 
\end{equation}
with the boundary conditions $\varphi_i (r = \infty) = 0$ and $\mathrm{d} \varphi_i  / \mathrm{d} r |_{r=0} = 0$, where $r$ is the spatial radial coordinate and $V$ is the effective potential at finite temperatures given in Eq.~(\ref{eq: Veff_Tneq0}). 
These solutions describe the profiles of the critical bubble. 

Once the solutions are obtained, $S_3$ is given by the following integral; 
\begin{equation}
S_3 = 4 \pi \int_0^\infty \mathrm{d}r \, 
r^2 \biggl[
	\sum_{i=1}^3 \frac{ 1 }{ 2 } 
		\left( 
		\frac{ \mathrm{d} \varphi_i }{ \mathrm{d} r }
		\right)^2
	+ V(\varphi_1, \varphi_2, \varphi_3; T )
\biggr]. 
\end{equation}
The nucleation temperature $T_n$ is given by $\Gamma / H^4 = 1$, where $H$ is the Hubble parameter.
It can be roughly estimated by $S_3 / T|_{T = T_n} = 140$~\cite{Grojean:2006bp}.
In the following, the solutions of Eq.~(\ref{eq: bubble_equation}) at $T_n$ are denoted by $\hat{\varphi}_1(z)$, $\hat{\varphi}_2(z)$, and $\hat{\varphi}_3(z)$.

\subsection{Numerical evaluations}

In this subsection, we show some numerical evaluations of the electroweak phase transition in the model. 
For simplicity, we consider only the single-step phase transition.  
For numerical evaluation, we used \texttt{CosmoTransitions}~\cite{Wainwright:2011kj}, which is a set of Python modules for calculating the effective potential and the Euclidian action. 

\begin{figure}[t]
    \centering
    \includegraphics[width=0.49\linewidth]{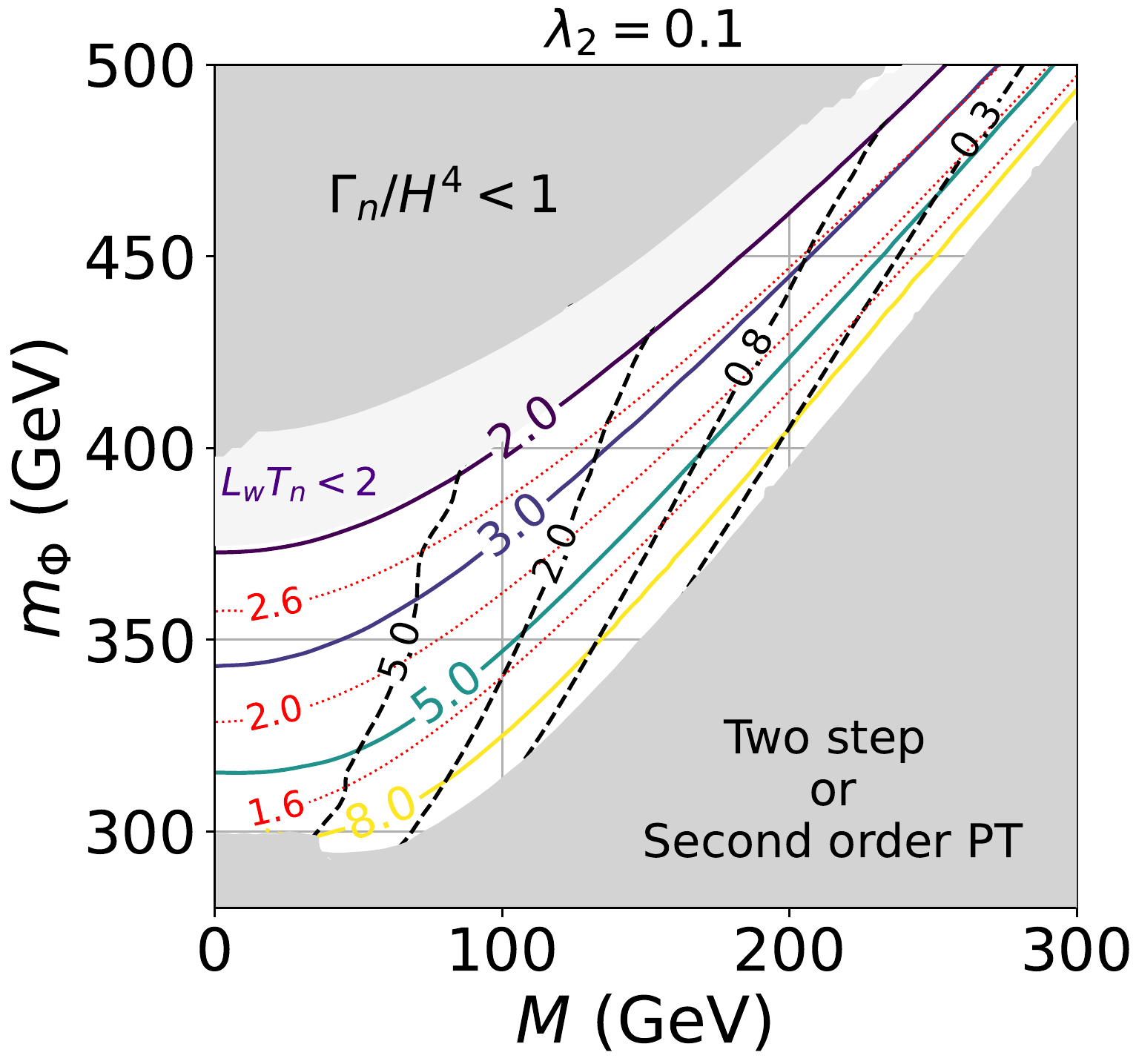}
    \includegraphics[width=0.49\linewidth]{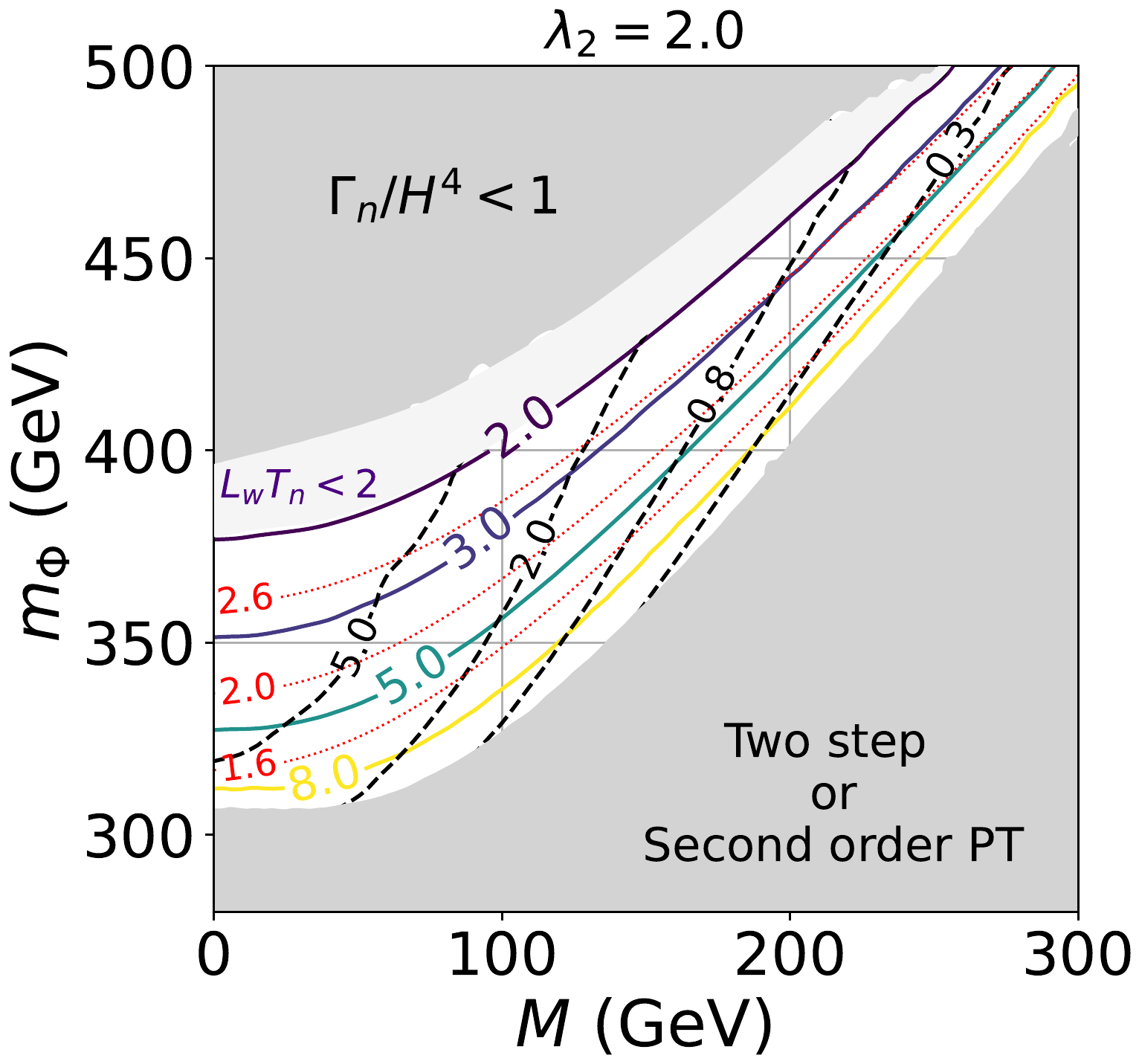}
    \caption{The behavior of the electroweak phase transition for various masses of the additional Higgs bosons and the decoupling parameter $M$. The masses of the additional Higgs bosons are assumed to be degenerated and given by $m_{\Phi}$. The red dotted lines are contours for $v_n/T_n$, and the black dashed lines are $10 \times \partial_z \theta|_\mathrm{max}$. Other colored lines are contours for $L_w T_n$. In the left (right) figure, $\lambda_2$ is set to be 0.1 (2.0). }
    \label{fig:mvsM}
\end{figure}

In figure~\ref{fig:mvsM}, we show the behavior of the electroweak phase transition for various masses of the additional Higgs bosons and the decoupling parameter $M$. Here, we assume that the additional Higgs bosons have the same mass $m_{\Phi}$. 
We show the figures for $\lambda_2 = 0.1$ (left) and $\lambda_2 = 2.0$ (right). 
Other parameters of the effective potential are set to be as follows; 
\begin{equation}
|\zeta_u| = 0.15, \quad \theta_u = -2.7, \quad 
|\lambda_7| = 0.8, \quad \theta_7 = -0.9. 
\end{equation}

In the lower gray region, the electroweak phase transition is two step or second-order. 
We do not consider this region. 
In the upper gray region, the nucleation rate per Hubble volume $\Gamma_n / H^4$ is less than 1. 
In such a region, the electroweak phase transition is not completed until the present. We do not thus consider this region. 

In the white regions, the electroweak phase transition is of the first-order, and occurs in a single step.
For successful electroweak baryogenesis, 
the electroweak phase transition has to be strongly first-order, where the sphaleron transition decouples inside the bubble quickly enough. 
The condition for realizing this situation is called the sphaleron decoupling condition, and it is roughly evaluated as $v_n / T_n \gtrsim 1$, where $v_n$ is the VEV at $T_n$~\cite{Kuzmin:1985mm}. 
In figure~\ref{fig:mvsM}, contours for $v_n/T_n$ are shown by the red lines. 
For a fixed value of $M$, the heavier $m_{\Phi}$ gives the larger $v_n/T_n$ because of the non-decoupling effect of the additional Higgs bosons~\cite{Kanemura:2004ch, Enomoto:2021dkl}. 
The minimum value of $v_n / T_n$ in the white regions is approximately 1.2, so that we can see that the sphaleron decoupling condition is satisfied in all the white regions.

Another important parameter for the strength of the electroweak phase transition is the wall width of the bubble $L_w$. 
For the stronger phase transition, $L_w$ is smaller. 
We evaluate $L_w$ by fitting the profile of the VEV $v(z) = \sqrt{ \hat{\varphi}_1^2 + \hat{\varphi}_2^2 +\hat{\varphi}_3^2}$ with the function
\begin{equation}
v(z) =  \frac{ v_n }{ 2 } \left( 1 - \mathrm{tanh} \frac{ z }{ L_w} \right), 
\end{equation}
where $z$ is the radial coordinate in the wall frame, where the bubble wall is stationary at $z=0$. 
In figure~\ref{fig:mvsM}, contours for $L_w T_n = 2.0$, $3.0$, $5.0$, and $8.0$ are shown in purple, dark blue, dark green, and yellow lines, respectively. 
For a fixed value of $M$, $L_w T_n$ is smaller for the heavier additional Higgs bosons. 

For producing baryon asymmetry, the phase of the local mass of the top quark is important as we will discuss later \cite{Cline:2000nw}.
It is defined as 
\begin{equation}
\mathcal{L}_{\mathrm{mass}} = -m_t(z) e^{i \theta(z)} \overline{t_L} t_R + \mathrm{h.c.},
\end{equation}
where $m_t(z)$ and $\theta(z)$ are the absolute value and the phase of the local mass of the top quark, respectively.
$m_t(z)$ and $\partial \theta(z) / \partial z$ are given by~\cite{Cline:2011mm} \footnote{The sign of the second term in the right hand side of Eq.~(\ref{eq:localmass}) is opposite to that in Ref.~\cite{Cline:2011mm}.}
\begin{align}
    m_t (z) &= \frac{m_t}{v}(\hat{\varphi}_1^2 + 2|\zeta_u| \hat{\varphi}_1 \hat{\varphi}_H \cos (\theta_H + \theta_u) + |\zeta_u|^2 \hat{\varphi}_H^2 )^{1/2},  \\[6pt]
    \partial_z \theta (z) &\equiv \frac{\partial \theta}{\partial z} = -\frac{\hat{\varphi}_H^2}{\hat{\varphi}_1^2 + \hat{\varphi}_H^2} \partial_z \theta_H - \partial_z \tan^{-1} \left( \frac{\hat{\varphi}_H|\zeta_u|\sin(\theta_H + \theta_u)}{\hat{\varphi}_1 + \hat{\varphi}_H |\zeta_u| \cos(\theta_H + \theta_u)}  \right),
    \label{eq:localmass}
\end{align} 
where $\hat{\varphi}_H$ and $\theta_H$ are defined as
\begin{equation}
\label{eq: varphiH_thetaH}
\hat{\varphi}_H = \sqrt{ \hat{\varphi}_2^2 + \hat{\varphi}_3^2 }, \quad 
\theta_H = \mathrm{arctan}\left( \frac{ \hat{\varphi}_3 }{ \hat{\varphi}_2 } \right). 
\end{equation}
The spacial variation of $\theta(z)$ provides the source of CP-violation.
In figure~\ref{fig:mvsM}, contours for the maximal value of $\partial_z \theta$ are shown: $10 \times \partial_z \theta |_\text{max} = 0.3$, $0.8$, $2.0$, and $5.0$ in $\mathrm{GeV}^{-1}$. 
We can see that the maximum value $\partial_z \theta |_\text{max}$ decreases as the decoupling parameter $M$ increases. 
In addition, it is smaller for the larger $\lambda_2$. 
This behavior can be understood as follows. 
The parameter $M^2$ and $\lambda_2$ is the coefficient of $|\Phi_2|^2$ and $|\Phi_2|^4$ in the Higgs potential. 
The potential is thus higher for the configurations $\varphi_2$ and $\varphi_3$ for larger $M^2$ and $\lambda_2$. 
The bubble profile $\hat{\varphi}_2$ and $\hat{\varphi}_3$ then cannot be far away from $\hat{\varphi}_2(z) = 0$ and $\hat{\varphi}_3(z) = 0$, respectively. 
In summary, $\partial_z \theta |_\text{max}$ is smaller for larger $M^2$ and $\lambda_2$.

\section{Baryon asymmetry of the universe \label{sec:BAU}}

\subsection{Transport equations and baryon asymmetry}
\label{subsec:TRANS}

In this subsection, we show the transport equation for the charge transport of the top quark in the WKB method~\cite{Joyce:1994fu, Joyce:1994zn, Cline:2000nw, Fromme:2006wx, Cline:2020jre}. 
According to ref.~\cite{Cline:2020jre}, we consider the transport equation including the relativistic effect of the wall velocity $v_w$. 

In the following, $v_w$ is assumed to be a constant. 
We discuss the problem in the wall frame, where the bubble wall is stationary. 
The radial direction in the wall frame is denoted by $z$. 
The bubble wall is located at $z=0$. 
The positive (negative) direction of $z$ is the symmetric (broken) phase. 

In the WKB method~\cite{Cline:2000nw, Fromme:2006wx}, 
the group velocity $v_g$ and the semi-classical force $F$ for the WKB state of the particle are given by 
\begin{equation}
\begin{array}{l}
\displaystyle{v_g = \frac{p_z}{E}\left( 1 \pm s\frac{m^2 \theta^\prime}{2E^2 E_z} \right)}, \\[20pt]
\displaystyle{F = -\frac{(m^2)^\prime}{2E} \pm s\frac{(m^2\theta^\prime)^\prime}{2EE_z}\mp s\frac{m^2 \theta^\prime (m^2)^\prime}{4E^3 E_z}}, \\
\end{array}
\label{TRANSforce}
\end{equation}
by solving the Dirac equation with the local mass $m$. 
The absolute value and the phase of the local mass of the particle are denoted by $m$ and $\theta$ in Eq.~(\ref{TRANSforce}), respectively.
The upper (lower) signs correspond to the particle (anti-particle).
The spin and the energy of the particle are denoted by $s$ and $E$, respectively. 
The symbol $E_z$ is defined as $E_z = \sqrt{p_z^2 + m^2}$, where $p_z$ is the kinetic momentum along the $z$ axis. 
The prime ${}^\prime$ in Eq.~(\ref{TRANSforce}) denotes the derivative by $z$. 

By using $v_g$ and $F$, 
the Boltzmann equation for the distribution function of the particle labeled by $i$ is given by  
\begin{equation}
    ( v_g \partial_{z} + F \partial_z ) f_i = C_i[f_i,f_j,\dots], 
    \label{TRANSboltzman}
\end{equation}
where $C_i[f_i,f_j,\dots]$ is the collision term. 
By assuming that the deviation from the thermal equilibrium is a small perturbation, 
the distribution function $f_i$ is expanded as~\cite{Fromme:2006wx}
\begin{equation}
f_i \simeq f_{0w} \pm \gamma s f_{0w}^\prime \Delta E - f_{0w}^\prime \mu_i \mp \gamma s f_{0w}^{\prime \prime} \Delta E \mu_i + \frac{1}{2}f_{0w}^{\prime \prime}(\gamma \Delta E)^2 + \delta f_i,
\end{equation}
where $f_{0w}$ is the thermal equilibrium distribution in the wall frame; 
\begin{equation}
f_{0w} = \frac{1}{e^{\beta[\gamma(E + v_w p_z)]} \pm 1}, \quad 
(-\text{ for bosons},\ +\text{ for fermions}). 
\end{equation}
The Lorentz factor is denoted by $\gamma = 1/ \sqrt{ 1 - v_w^2 }$. 
The functions $f_{0 w}^\prime$ and $f_{0 w}^{\prime \prime}$ are 
the derivative and the second derivative of $f_{0 w}$ by $\gamma E$.
$\Delta E$ is the difference of the energy between the particle $i$ and its anti-particle $\overline{i}$, which is given by
\begin{equation}
\Delta E = - \frac{m^2 \theta^{\prime}}{2EE_z}.
\end{equation}
The perturbation $\delta f_i$ and the chemical potential $\mu_i$ describe the deviation from the kinetic and the chemical equilibrium, respectively. 

By taking the difference between the Boltzmann equations for the particle and its anti-particle, 
the equation for the CP-odd deviations $\mu_{o,i} \equiv (\mu_i - \mu_{\overline{i}}) / 2$ and $\delta f_{o,i} = (\delta f_i - \delta f_{\overline{i}} )/2$ is given by~\cite{Fromme:2006wx, Cline:2020jre} 
\begin{align}
   -\frac{p_z}{E}f_{0w}^\prime \mu_{o,i}^\prime &+ \gamma v_w \frac{(m^2)^\prime}{2E}f_{0w}^{\prime\prime}\mu_{o,i} +\frac{p_z}{E} \partial_z \delta f_{o,i} - \frac{(m^2)^\prime}{2E}\partial_{p_z}\delta f_{o,i} \notag \\
    &+ s \gamma v_w \frac{(m^2 \theta^\prime)^\prime}{2EE_z}f_{0w}^\prime + s \gamma v_w \frac{m^2 \theta^\prime (m^2)^\prime}{4E^2 E_z}\left(\gamma f_{0w}^{\prime\prime} - \frac{f_{0w}^\prime}{E} \right) = C_o,
    \label{eq:trans1} 
\end{align}
where $C_o = (C_i - C_{\overline{i}})/2$. 
The fifth and sixth terms of the left-hand side of the equation are the source terms to produce the CP asymmetry. 

In order to eliminate the momentum variables from Eq.~(\ref{eq:trans1}), we integrate the both sides of Eq.~(\ref{eq:trans1}) over three-dimensional momentum, weighting by 1 and by $p_z / E$.
As a result, the transport equations for the chemical potential $\mu_i$ and the plasma velocity in the wall frame which is defined by $u_i = \int d^3 p(p_z / E) \delta f_i$ are given by~\cite{Cline:2020jre}
\begin{equation}
\label{eq: transport_eq}
\left\{
\begin{array}{l}
-D_{1i} \mu_i^\prime +u_{i}^\prime + \gamma v_w (m_i^2)^\prime Q_{1i} \mu_i - K_{0i} \overline{\Gamma}_i = h S_{1i}.  \\
-D_{2i} \mu_i^\prime -v_w u_{i}^\prime + \gamma v_w (m_i^2)^\prime Q_{2i} \mu_i + (m_i^2)^\prime \overline{R}_i u_i + \Gamma_{i,\mathrm{tot}} u_i + v_w K_{0i} \overline{\Gamma}_i = h S_{2i}, \\
\end{array}
\right.
\end{equation}
Here, the subscript $o$ is omitted. 
The source term $S_{li}$ is defined as 
\begin{equation}
    S_{li} = -\gamma v_w (m_i^2 \theta_i^\prime)^\prime Q_{li}^8 + \gamma v_w m_i^2 \theta_i^\prime(m_i^2)^\prime Q_{li}^9, \quad (l=1,2). 
\end{equation}
The functions $D$, $Q$, $\overline{R}$, and $K_0$ are defined in ref.~\cite{Cline:2020jre}.\footnote{The formula for $K_0$ in ref.~\cite{Cline:2020jre} includes an error as indicated in ref.~\cite{Lewicki:2021pgr}.}
In the derivation of Eq.~(\ref{eq: transport_eq}), we use the approximation that the spin $s$ is evaluated as $s = h \, \mathrm{sign}(p_z)$, where $h$ is the helicity of the particle. 
The symbol $\Gamma_{i,\mathrm{tot}}$ denotes the total reaction rate of the particle $i$, 
and $\overline{\Gamma}_i$ is the sum of the reaction rate for inelastic scattering processes including $i$. 

In the following, we consider the transport equations in the aligned THDM. 
We neglect the fermion masses except for the top quarks. 
For inelastic scattering processes, 
the strong sphaleron process, the $W$ boson scattering, the top Yukawa interaction, the top helicity flips and the Higgs number violation are considered. 
The reaction rates for each inelastic process are denoted by $\Gamma_{ss}, \Gamma_W, \Gamma_y, \Gamma_m$ and $\Gamma_h$, respectively. 
We refer to their values in ref.~\cite{Cline:2020jre}. 

The transport equation for right-handed bottom quarks and quarks in the first and second generations can be analytically solved. 
Their chemical potentials are then represented by the linear combination of the chemical potentials for top quarks and left-handed bottom quarks. 
By substituting these solutions, the transport equations for top quarks, the left-handed bottom quarks and the Higgs doublets are given as follows~\cite{Cline:2020jre}; 
\begin{itemize}

\item Left-handed top quarks ($t$)
\begin{equation}
\left\{
\begin{array}{l}
-D_{1t} \mu_t^\prime + u_t^\prime + \gamma v_w (m_t^2)^\prime Q_{2t} \mu_t 
    - K_{0t}  \overline{\Gamma}_t  = -S_{1t}, \\
-D_{2t} \mu_{t}^\prime -  v_w u_{t}^\prime + \gamma v_w (m_t^2)^\prime Q_{t2} \mu_t + (m_t^2)^\prime \overline{R}_t u_{t} + \Gamma_{t,\mathrm{tot}} u_t + v_w K_{0t}  \overline{\Gamma}_t  = -S_{2t}. \\
\end{array}
\right.
\end{equation}

\item Left-handed bottom quarks ($b$)
\begin{equation}
\left\{
\begin{array}{l}
-D_{1b} \mu_b^\prime + u_b^\prime
    - K_{0b}  \overline{\Gamma}_b  = 0, \\
-D_{2b} \mu_b^\prime -  v_w u_b^\prime + \Gamma_{b,\mathrm{tot}} u_b + v_w K_{0b}  \overline{\Gamma}_b  = 0. \\
\end{array}
\right.
\end{equation}

\item Charge conjugation of right-handed singlet top quarks ($t^c$)
\begin{equation}
\left\{
\begin{array}{l}
-D_{1t} \mu_{t^c}^\prime + u_{t^c}^\prime + \gamma v_w (m_t^2)^\prime Q_{2t} \mu_{t^c} 
    - K_{0t}  \overline{\Gamma}_{t^c}  = -S_{1t}, \\
-D_{2t} \mu_{t^c}^\prime -  v_w u_{t^c}^\prime + \gamma v_w (m_t^2)^\prime Q_{t2} \mu_{t^c} + (m_t^2)^\prime \overline{R}_t u_{t^c} + \Gamma_{t,\mathrm{tot}} u_{t^c} + v_w K_{0t}  \overline{\Gamma}_{t^c}  = -S_{2t}. \\
\end{array}
\right.
\end{equation}

\item The Higgs doublets ($h$)
\begin{equation}
\left\{
\begin{array}{l}
-D_{1h} \mu_h^\prime + u_h^\prime
    - K_{0h}  \overline{\Gamma}_h  = 0, \\
-D_{2h} \mu_h^\prime -  v_w u_h^\prime + \Gamma_{h,\mathrm{tot}} u_h + v_w K_{0h} \overline{\Gamma}_h  = 0. \\
\end{array}
\right.
\end{equation}

\end{itemize}
The inelastic reaction rates for each particle is defined as~\cite{Cline:2020jre}
\begin{align}
    \overline{\Gamma}_t &= \Gamma_{ss} \Bigl((1 + 9D_{0t}) \mu_t + 10 \mu_b + (1-9D_{0t}) \mu_{t^c} \Bigr)  \notag \\
    &~~+ \Gamma_W (\mu_t - \mu_b) + \Gamma_y (\mu_t + \mu_{t^c} + \mu_h) +2 \Gamma_m (\mu_t + \mu_{t^c}),  \\
    \overline{\Gamma}_b &= \Gamma_{ss} \Bigl( (1 + 9D_{0t}) \mu_t + 10 \mu_b + (1 + 9D_{0t}) \mu_{t^c} \Bigr) \notag \\
    &~~+\Gamma_W (\mu_b - \mu_t) + \Gamma_y (\mu_b + \mu_{t^c} + \mu_h),  \\
    \overline{\Gamma}_{t^c} &= \Gamma_{ss} \Bigl( (1 + 9D_{0t}) \mu_t + 10 \mu_b + (1 - 9D_{0t}) \mu_{t^c} \Bigr) \notag \\
    &~~+ 2 \Gamma_m ( \mu_{t^c} + \mu_t) + \Gamma_y (2 \mu_{t^c} + \mu_t + \mu_b + 2 \mu_h),  \\
    \overline{\Gamma}_h &= \frac{3}{4} \Gamma_y (2 \mu_h + \mu_t + \mu_b + 2\mu_{t^c}) + \Gamma_h \mu_h.
\end{align}
$m_t$ and $\partial_z \theta_t$ are given in Eqs. (\ref{eq:localmass}).

By solving the above transport equations, 
the distributions of the chemical potentials are obtained. 
By using these distributions, the produced baryon number density normalized by the entropy density can be evaluated as~\cite{Cline:2000nw, Cline:2011mm}
\begin{equation}
    \eta_B = \frac{405 \Gamma_{\mathrm{sph}}}{4 \pi^2 v_w g_\ast T_n} \int_0^\infty dz ~ \mu_{B_L} f_{\mathrm{sph}} (z) \, \mathrm{exp}\left( -\frac{45 \Gamma_{\mathrm{sph}} z}{4 v_w} \right),
\end{equation}
where $\mu_{B_L}$ is defined as 
\begin{equation}
    \mu_{B_L} = \frac{1}{2} (1 + 4D_{0t})\mu_t + \frac{1}{2}(1 + 4D_{0b}) \mu_b - 2 D_{0t} \mu_{t^c}. 
\end{equation}
The symbol $g_\ast$ is the effective degree of freedom for the entropy. 
The weak sphaleron rate in the symmetric phase is denoted by $\Gamma_\mathrm{sph}$. 
By the lattice calculations, $\Gamma_\mathrm{sph}$ is evaluated as $\Gamma_\mathrm{sph} = 1.0 \times 10^{-6} T$~\cite{Moore:2000mx}. 
The function $f_\mathrm{sph}(z)$ describes the suppression of the weak sphaleron rate outside the bubble caused by the nonzero VEV. 
According to ref.~\cite{Cline:2011mm}, we evaluate $f_\mathrm{sph}$ as 
\begin{equation}
f_{\mathrm{sph}}(z) = \mathrm{min}
\Bigl\{1, \ \frac{2.4 T}{\Gamma_{\mathrm{sph}}} e^{-\frac{40 v_n(z)}{T}} \Bigr\}. 
\end{equation}

\subsection{Numerical results for the BAU}
In this subsection, we show the numerical evaluations for the BAU in the model.
Although the wall velocity $v_w$ is an important parameter for the calculation of the baryon density, we treat it as a free parameter, following to the former discussions \cite{Cline:2000nw, Fromme:2006cm, Fromme:2006wx, Cline:2011mm, Cline:2020jre}.

\begin{figure}[t]
    \centering
    \includegraphics[width=0.7\linewidth]{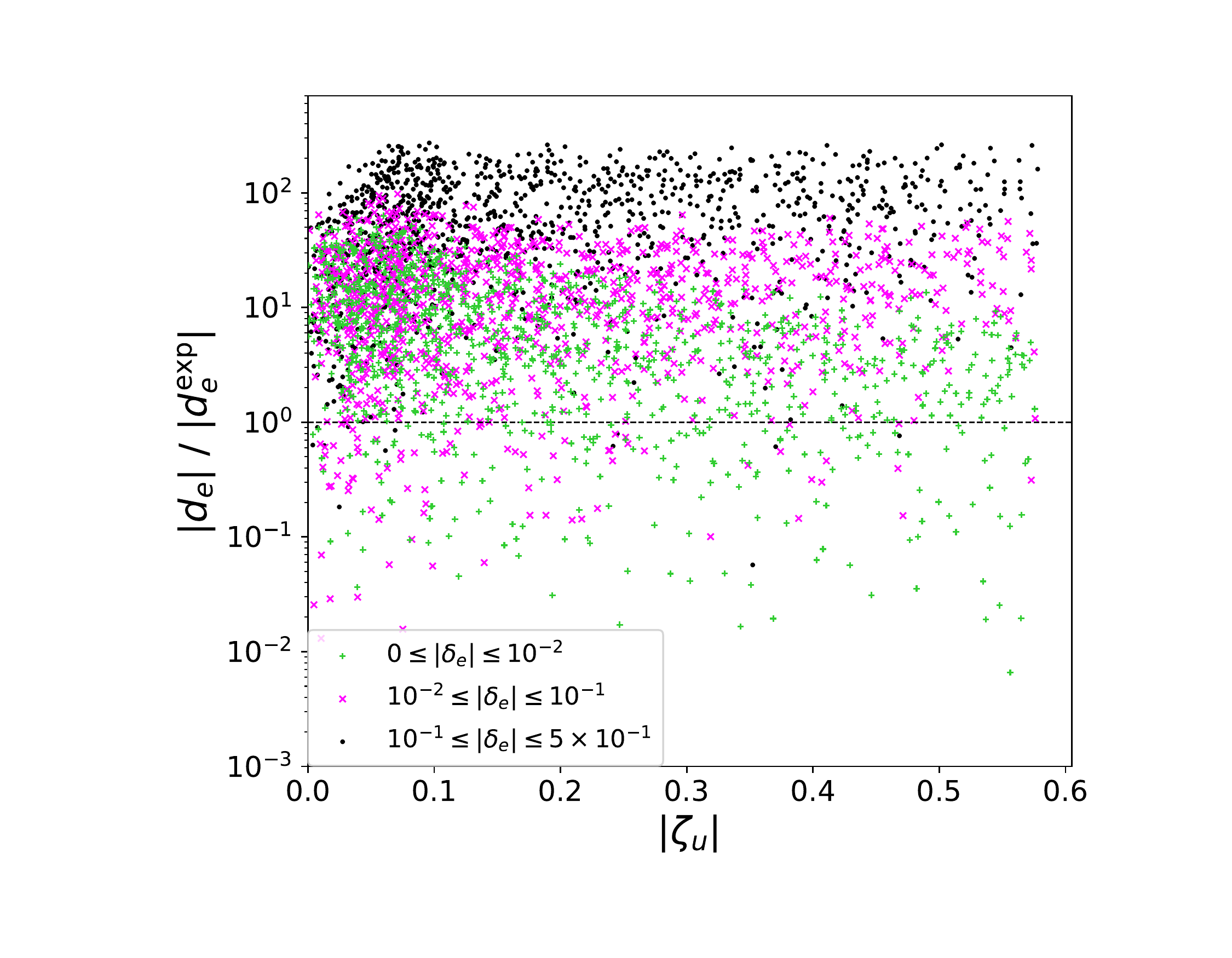}
    \caption{The correlations between $|\zeta_u|$ and $|d_e| / |d_e^{\mathrm{exp}}|$. Input parameters are given by $\lambda_2 = 0.1, ~m_{\Phi} = 350~\mathrm{GeV},~M = 30~\mathrm{GeV}$ and  $v_w = 0.1$. We scan the other parameters for the regions of $\theta_u = \theta_d = [0,2\pi),~|\zeta_u| = [0,0.6],~|\zeta_d|=|\zeta_e|=[0,10],~|\lambda_7|=[0.5,1.0]$ and $\theta_7 = [0,2\pi)$. Black, magenta and green points are the cases of $0.1 \le  |\delta_e| \le 0.5,~ 0.01 \le |\delta_e| \le 0.1$ and $0 \le |\delta_e| \le 0.01$, respectively. Here, we have defined $\delta_e \equiv \theta_u - \theta_e$.}
    \label{fig:scat1}
\end{figure}
\begin{figure}[t]
    \centering
    \includegraphics[width=0.55\linewidth]{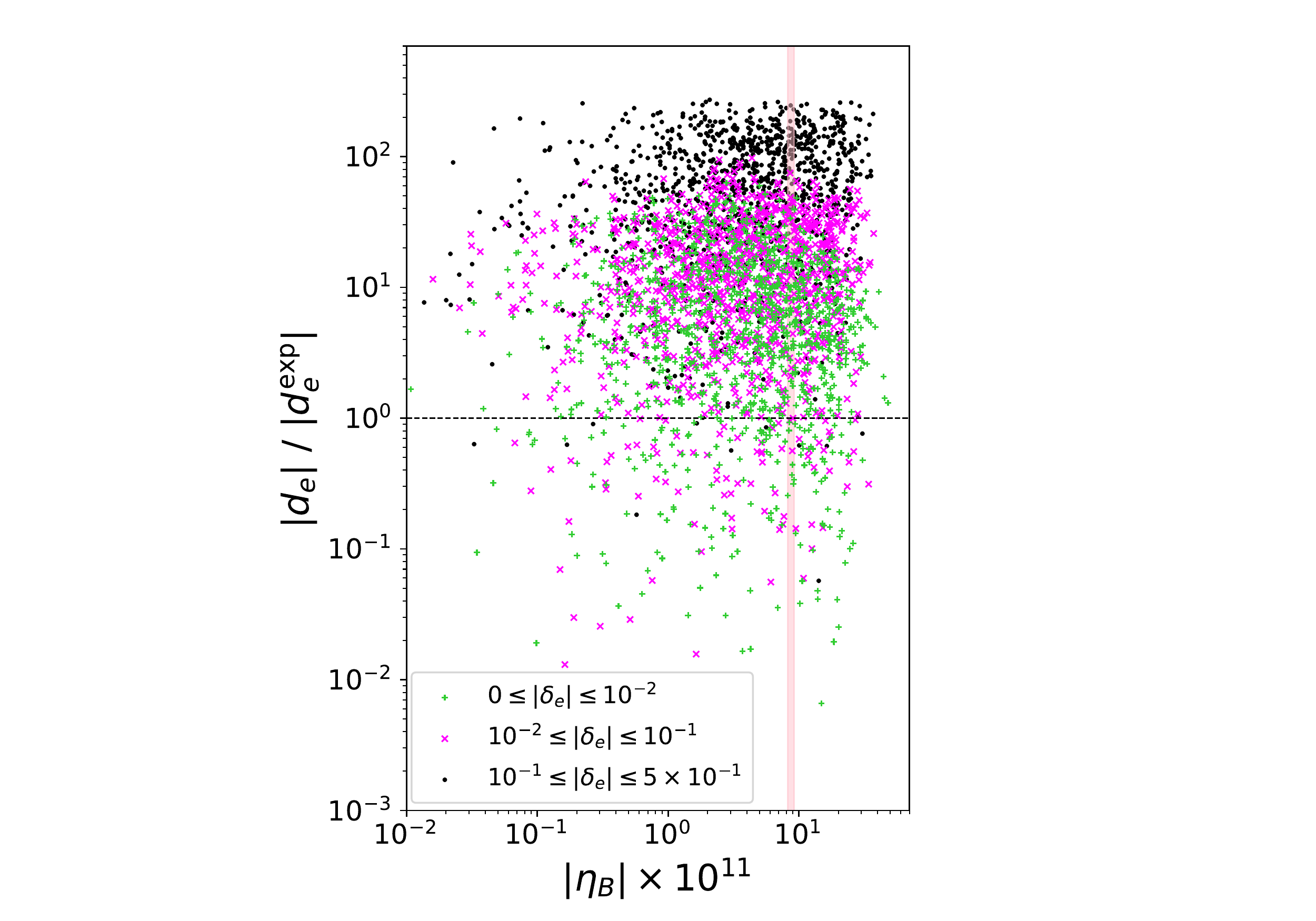}
    \caption{The correlation between $|d_e| / |d_e^{\mathrm{exp}}|$ and $|\eta_B| \times 10^{11}$. Input parameters are the same as in figure~\ref{fig:scat1}. The vertically parallel pink region explains the observed BAU shown in Eq.~(\ref{eq:OBSBAU}) within 95\%~C.L.}
    \label{fig:scat2}
\end{figure}
In figure~\ref{fig:scat1}, the correlation between $|\zeta_u|$ and $|d_e| / |d_e^{\mathrm{exp}}|$, which is the eEDM normalized by the current experimental bound $|d_e^{\mathrm{exp}}| = 1.0 \times 10^{-29}~e$ cm, is shown.
Input parameters are $\lambda_2 = 0.1, ~M = 30~\mathrm{GeV}, ~m_{\Phi} = 350~\mathrm{GeV}, ~ v_w = 0.1$, and we scan the other parameters for the regions of $\theta_u = \theta_d = [0,2\pi),~ |\zeta_u| = [0,0.6], ~|\zeta_d|=|\zeta_e|=[0,10],~|\lambda_7|=[0.5,1.0]$ and $\theta_7 = [0,2\pi)$.
Black, magenta, and green points are the cases of $0.1 \le  |\delta_e| \le 0.5,~ 0.01 \le |\delta_e| \le 0.1, ~ 0 \le |\delta_e| \le 0.01$, respectively.
Here, we have defined $\delta_e \equiv \theta_u - \theta_e$.
Each point is allowed by the theoretical and current experimental constraints except for the eEDM data, which have been discussed in section~\ref{sec:CONST}.
The black dashed line in figure~\ref{fig:scat1} is the current experimental bound of the eEDM, and the points above this line have been excluded. 
As we mentioned in section~\ref{sec:CONST}, the fermion and scalar loop diagrams which contribute to the eEDM are approximately proportional to  $|\zeta_u||\zeta_e|\sin \delta_e$ and $|\zeta_e||\lambda_7|\sin(\theta_7 - \theta_e)$, respectively.
When $|\zeta_u| \lesssim 0.1$, the fermion loop contributions are small, so that $|\delta_e|$ dependence of the eEDM shown in figure~\ref{fig:scat1} is small.
On the other hand, in the region $|\zeta_u| \gtrsim 0.1$, the eEDM becomes large as $|\delta_e|$ increases.
Some benchmark points in figure~\ref{fig:scat1} are allowed from the eEDM data due to the destructive interference between the CP-violating effects in Barr--Zee type diagrams.

In figure~\ref{fig:scat2}, the correlation between $|d_e| / |d_e^{\mathrm{exp}}|$ and the absolute value of the normalized baryon density $|\eta_B| \times 10^{11}$ is shown.
Input parameters are the same as in figure~\ref{fig:scat1}.
The vertically parallel pink region explains the observed BAU with 95 \% C.L.
The eEDM decreases in the order of black, magenta and green, so that a lot of magenta and green points can generate sufficient BAU under the eEDM constraints.

\begin{figure}[t]
    \centering
    \includegraphics[width=0.9\linewidth]{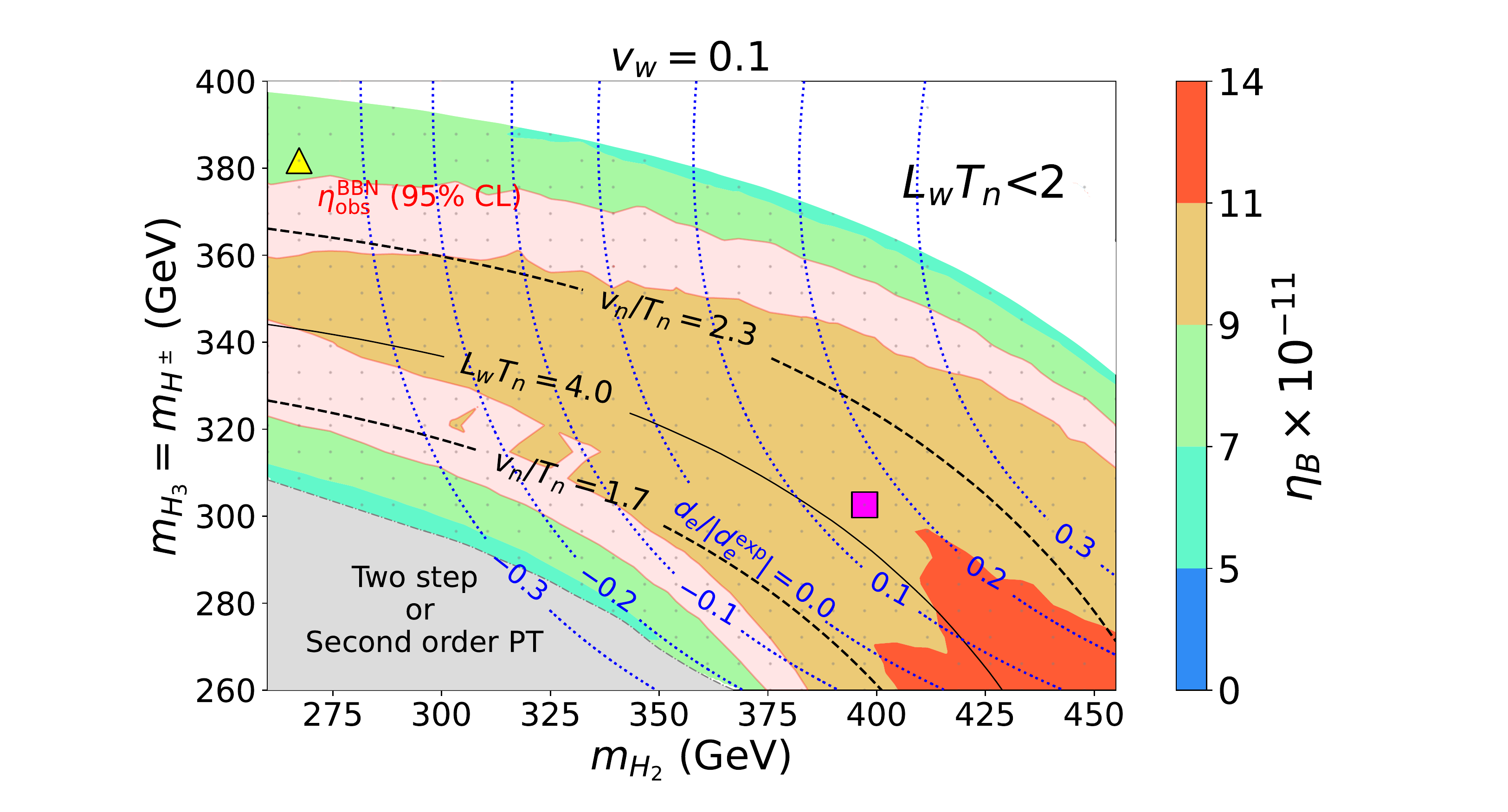}
    \includegraphics[width=0.9\linewidth]{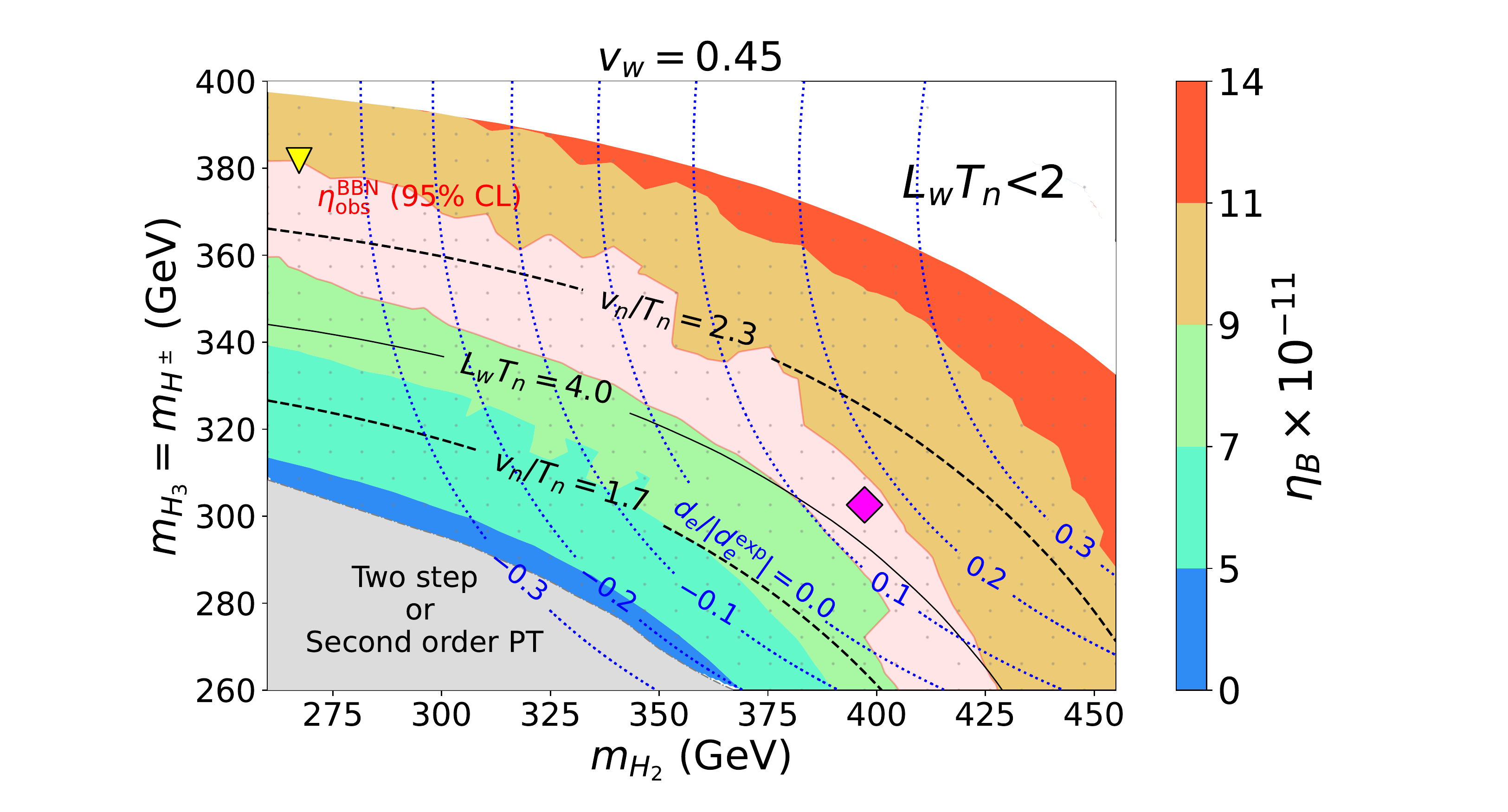}
    \caption{The generated BAU on the plane of $m_{H_2}$ and $m_{H_3}=m_{H^\pm}$ in the case of $v_w = 0.1$ (top) and $v_w = 0.45$ (bottom). Input parameters are shown in Eq. (\ref{eq:COLORINPUT}). The pink regions in each figure correspond to the observed BAU.}
    \label{fig:COLOR}
\end{figure}
Top (bottom) panel of figure \ref{fig:COLOR} shows generated BAU in the case of $v_w = 0.1$  (0.45) in the following benchmark point under the constraints: 
\begin{gather}
\label{eq:COLORINPUT}
M=30 ~\mathrm{GeV},~~   \lambda_2=0.1,  ~~ |\lambda_7|=0.8,  ~~ \theta_7=-0.9, \\ \notag
|\zeta_u|=|\zeta_d|=|\zeta_e|=0.18, ~~ \theta_u =\theta_d =-2.7,  ~~ \delta_e=-0.04.
\end{gather}
At the point in the gray region in figure~\ref{fig:COLOR}, the electroweak phase transition is two step or second order.
The black dashed lines in figure~\ref{fig:COLOR} are the contour of $v_n /T_n=1.7$ and $2.3$.
Therefore, the strongly first order phase transition for the sphaleron decoupling condition occurs above the gray region.
The black solid line is the contour of $L_w T_n =4.0$.
The white region satisfies $L_w T_n <2$, where the WKB approximation becomes invalid~\cite{Fromme:2006cm}.
As shown in figure~\ref{fig:mvsM}, since the invariant mass parameter $M$ is fixed, increasing the mass of heavy scalars makes the electroweak phase transition stronger, and makes $L_w T_n$ smaller.
In the case of $v_w = 0.1$, the generated BAU increases as the masses of the additional Higgs bosons increase up to $L_w T_n =4.0$, and then turns to decrease.
On the other hand, in the case of $v_w = 0.45$, the BAU gets larger as the phase transition is stronger.
In the both panels of figure~\ref{fig:COLOR}, the pink regions which are sandwiched by green and orange regions can explain the observed BAU with better than 95 \% C.L.

In the blue lines in figure \ref{fig:COLOR}, various values of the eEDM $d_e/|d_e^{\mathrm{exp}}|$ are shown.
The line of $d_e/|d_e^{\mathrm{exp}}| = 0.0$ in figure~\ref{fig:COLOR} is due to the destructive interference between two independent diagrams shown in figure~\ref{fig:BZdiagram}.
The upper bound of the eEDM is out of range of these panels, so that whole region is allowed by the eEDM experiment.
At future eEDM experiments, the upper limit is expected to be improved by an order of magnitude~\cite{ACME:2018yjb}.
In such a case, only the region within $|d_e|/|d_e^{\mathrm{exp}}| < 0.1$ will be allowed.

The nEDM in figure~\ref{fig:COLOR} is about four orders of magnitude smaller than the current upper bound~\cite{nEDM:2020crw} because of $\theta_u-\theta_d = 0$.
Even if $\theta_u-\theta_d \neq 0$, the nEDM is at most a half of the current limit.

\subsection{Phenomenological predictions for future experiments}
In this subsection, we discuss phenomenological predictions for future experiments.
We here set benchmark points which are colored with a yellow and magenta in each panel of figure~\ref{fig:COLOR}.
We define the upward triangle (downward triangle) point with yellow as BP1a (BP1b), and the square (diamond) point with magenta as BP2a (BP2b).
Magnitudes of the electroweak phase transitions $v_n / T_n$ in the BP1a and the BP1b are relatively stronger than the ones in the BP2a and the BP2b.
Table~\ref{tab:BPs} shows the input parameters of the four benchmark points $v_w$, $m_{H_{2}}$, $m_{H_3} (= m_{H^\pm})$ and $M$, as well as $v_n/T_n$, $L_w T_n$ and $\eta_B$ in these points are also shown.
BP1b and BP2b, which are the cases of $v_w = 0.45$, can explain the observed BAU from BBN in Eq. (\ref{eq:OBSBAU}).

First we discuss some testabilities for the strongly first order phase transition.
It is known that the loop effects of heavy Higgs bosons for the strongly first order phase transition increase $\Delta R$ which is the deviation of triple Higgs boson coupling from the SM value~\cite{Kanemura:2004ch}.
In Table~\ref{tab:BPs}, the values of $\Delta R$ at one loop level which is given by Eq.~(\ref{eq:TRIPLECOUPLING}) are shown in the second column from the last.
In the BP1 and BP2, the deviations of the triple Higgs boson coupling $\Delta R$ become 61\% and 44\%, respectively.
Therefore, these points would be tested at the HL-LHC~\cite{Cepeda:2019klc}, the future updated ILC~\cite{Fujii:2015jha, Bambade:2019fyw}, and CLIC~\cite{CLICdp:2018cto}.
\begin{table}[t]
\begin{center}
\begin{tabular}{|c|c|c|c|c|c|c|c|c|c|}
\hline
     & $v_w$   & $m_{H_2}$              & $m_{H_3,H^\pm}$              & $M$                 & $~v_n/T_n~$               & $~L_wT_n~$                & $\eta_B$ & $\Delta R$ & $\sigma \mathcal{B}(H_1 \to \gamma \gamma)$ \\ \hline
~BP1a~ &  ~0.1~   & \multirow{2}{*}{267 GeV} & \multirow{2}{*}{381 GeV} & \multirow{2}{*}{30 GeV} & \multirow{2}{*}{~2.4~} & \multirow{2}{*}{~2.6~} & $~7.8 \times 10^{-11}~$ & \multirow{2}{*}{~~0.61~~} &  \multirow{4}{*}{104 $\pm$ 5 fb}   \\ \cline{1-2} \cline{8-8} 
~BP1b~ & ~0.45~ &                   &                   &                   &                   &                   &  $9.1 \times 10^{-11}$ &  &  \\ \cline{1-9}  
~BP2a~ & ~0.1~  & \multirow{2}{*}{397 GeV} & \multirow{2}{*}{302 GeV} & \multirow{2}{*}{30 GeV} & \multirow{2}{*}{~2.0~} & \multirow{2}{*}{~4.1~} &  $~10.8 \times 10^{-11}~$ & \multirow{2}{*}{~~0.44~~} &  \\ \cline{1-2} \cline{8-8} 
~BP2b~ & ~0.45~ &                   &                   &                   &                   &                   &   $9.0 \times 10^{-11}$ &  & \\ \hline
\end{tabular}
\caption{The input parameters of the four benchmark points $v_w$, $m_{H_{2}}$, $m_{H_3} (= m_{H^\pm})$ and $M$ are shown. In addition, $v_n/T_n$, $L_w T_n$ and $\eta_B$ in each benchmark point are shown. The observed BAU shown in Eq.~(\ref{eq:OBSBAU}) can be explained in both BP1b and BP2b. As phenomenological predictions, we show the values of $\Delta R$ and $\sigma \mathcal{B}(H_1 \to \gamma\gamma)$ in these benchmark points. $\Delta R$ is the deviation in the triple Higgs boson coupling. $\sigma$ is the production cross section of the SM Higgs boson, and $\mathcal{B}(H_1 \to \gamma\gamma)$ is the branching ratio of the decay into a photon pair.}
\label{tab:BPs}
\end{center}
\end{table}

In addition to the triple Higgs boson coupling, the decay of the Higgs boson into a photon pair is affected by the non-decoupling effect of the charged Higgs boson \cite{HiggsGamma_Early1, HiggsGamma_Early2, Barroso:1999bf, Arhrib:2003vip, Djouadi:2005gj, Akeroyd:2007yh, Posch:2010hx}.
From the latest data at ATLAS Collaboration \cite{ATLAS:2020pvn}, the observed value of $\sigma \mathcal{B}(H_1 \to \gamma\gamma)$ is given by
\begin{equation}
\sigma \mathcal{B}(H_1 \to \gamma\gamma)_{\mathrm{obs}} = 127 \pm 10 ~ \mathrm{fb},
\end{equation}
where $\sigma$ is the production cross section of the SM Higgs boson and $\mathcal{B}(H_1 \to \gamma\gamma)$ is the branching ratio of the decay into di-photon.
The theoretical value of $\sigma \mathcal{B}(H_1 \to \gamma\gamma)$ in BP1 and BP2 is shown in the last column of Table \ref{tab:BPs}.
In BP1 and BP2, we obtain $\sigma \mathcal{B}(H_1 \to \gamma\gamma) = 104 \pm 5$~fb, and the uncertainty stems from theoretical errors of the production cross section of the SM Higgs boson.
Unlike the behavior of the non-decoupling effect in $\Delta R$ in Eq.~(\ref{eq:TRIPLECOUPLING}), the Higgs di-photon decay depends on a coupling proportional to $(m_{H^\pm}^2 - M^2)/m_{H^\pm}^2$. 
This effect does not decouple and becomes a constant for $m_{H^\pm} \to \infty$.
Therefore, the values of $\sigma \mathcal{B}(H_1 \to \gamma\gamma)$ in the BP1 and BP2 are the same within the range of significant figures.
The predictions on $\sigma \mathcal{B}(H_1 \to \gamma\gamma)$ in BP1 and BP2 overlap with the observed value within $2 \sigma$ significance.
These benchmark points would be tested by the precision measurement of the Higgs di-photon search at the future colliders such as the HL-LHC~\cite{Cepeda:2019klc}.

Furthermore, the GWs can also be produced from the collision of the bubbles which are created at the first order phase transition \cite{Grojean:2006bp, Caprini:2015zlo, Kakizaki:2015wua, Hashino:2016rvx, Espinosa:2010hh}. 
The sources of the GWs are composed by the contributions from the scalar field $\Omega_{\phi}(f)$, the sound waves of the plasma $\Omega_{\mathrm{sw}}(f)$, and the magnetohydrodynamics (MHD) turbulence $\Omega_{\mathrm{turb}}(f)$, where $f$ is the frequency of the GWs.
Each contribution is decided by the wall velocity $v_w$, the latent heats $\alpha$, and the time duration $\tilde{\beta}$ at the phase transition.
These can be defined by 
\begin{align}
	&\alpha = \rho_{\mathrm{vac}} (T_n)/ \rho_{\mathrm{rad}} (T_n), \notag \\
	&\tilde{\beta} = T_n (dS_3 / dT)_{T=T_n},
\end{align}
where,
\begin{align}
    &\rho_{\mathrm{vac}}(T_n) = \left(- \Delta V + T \frac{\partial \Delta V}{ \partial T} \right)_{T = T_n},  \notag \\
    &\rho_{\mathrm{rad}}(T_n) = g_* \pi^2 T_n^4 / 30,
\end{align}
where $\Delta V$ is the difference between the values of the effective potential of the true vacuum and the false vacuum.
The total energy density of the GWs is given by 
\begin{equation}
    h^2 \Omega_{\mathrm{GW}}(f) = h^2 \Omega_{\phi} +h^2 \Omega_{\mathrm{sw}} + h^2 \Omega_{\mathrm{turb}}.
\end{equation}
In THDMs, the terminal wall velocity does not reach the speed of light~\cite{Caprini:2015zlo}.
For simplicity, we assume that the velocity for the numerical analysis of the GWs matches the one used in the calculations of the BAU.
In this scenario, the leading contribution is the sound waves~\cite{Kakizaki:2015wua}, and the contribution from the scalar field is negligible.\footnote{The contribution from the turbulence is decided by a part of the latent heats $\epsilon \kappa_v\alpha$, where $\kappa_v$ is defined by how the latent heats are transformed into the bulk motion of the plasma fluid. In the following analysis, we set $\epsilon = 0.05$~\cite{Caprini:2015zlo, Hashino:2016rvx}.}

\begin{figure}[t]
    \centering
    \includegraphics[width=0.45\linewidth]{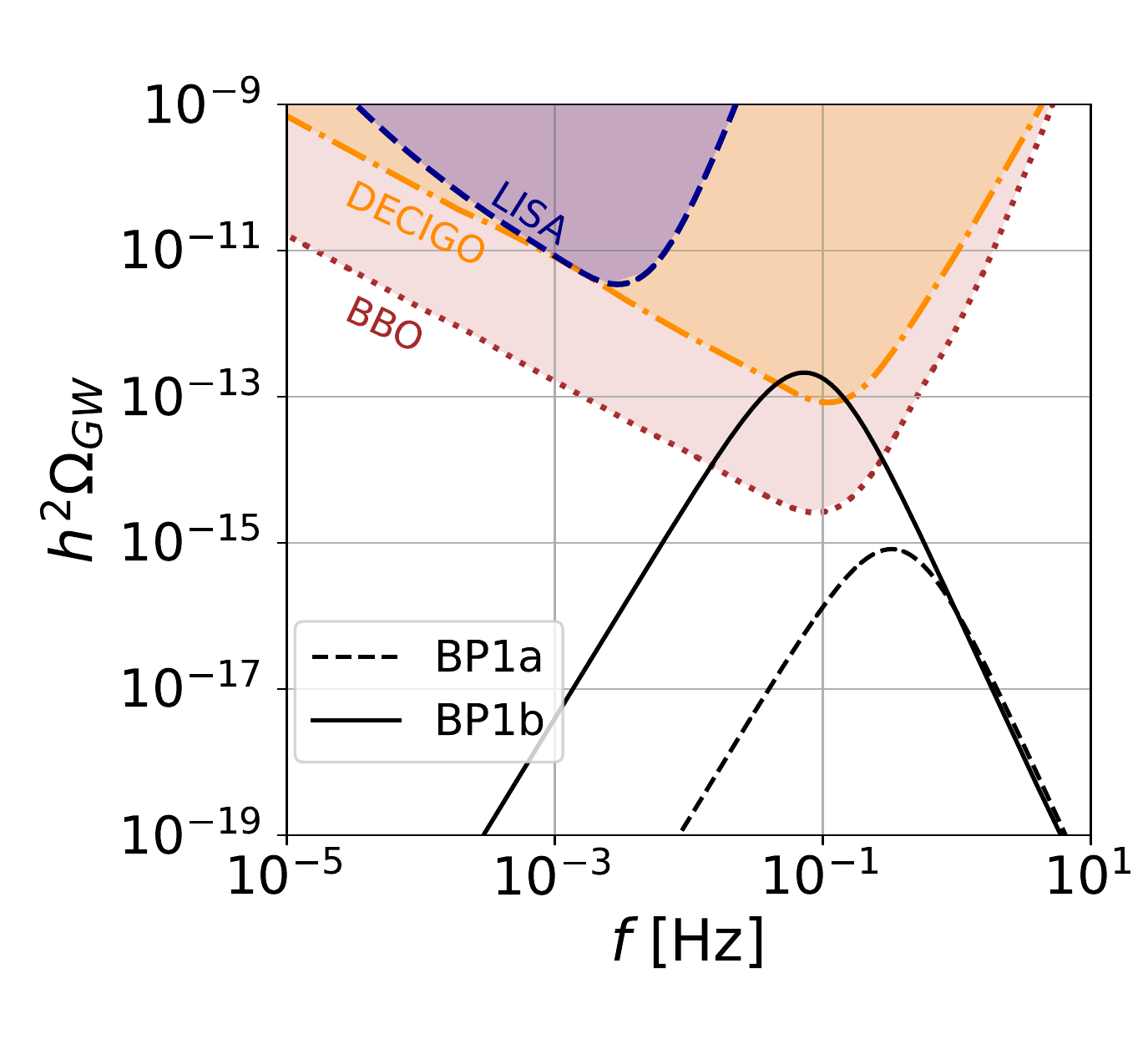}
    \includegraphics[width=0.45\linewidth]{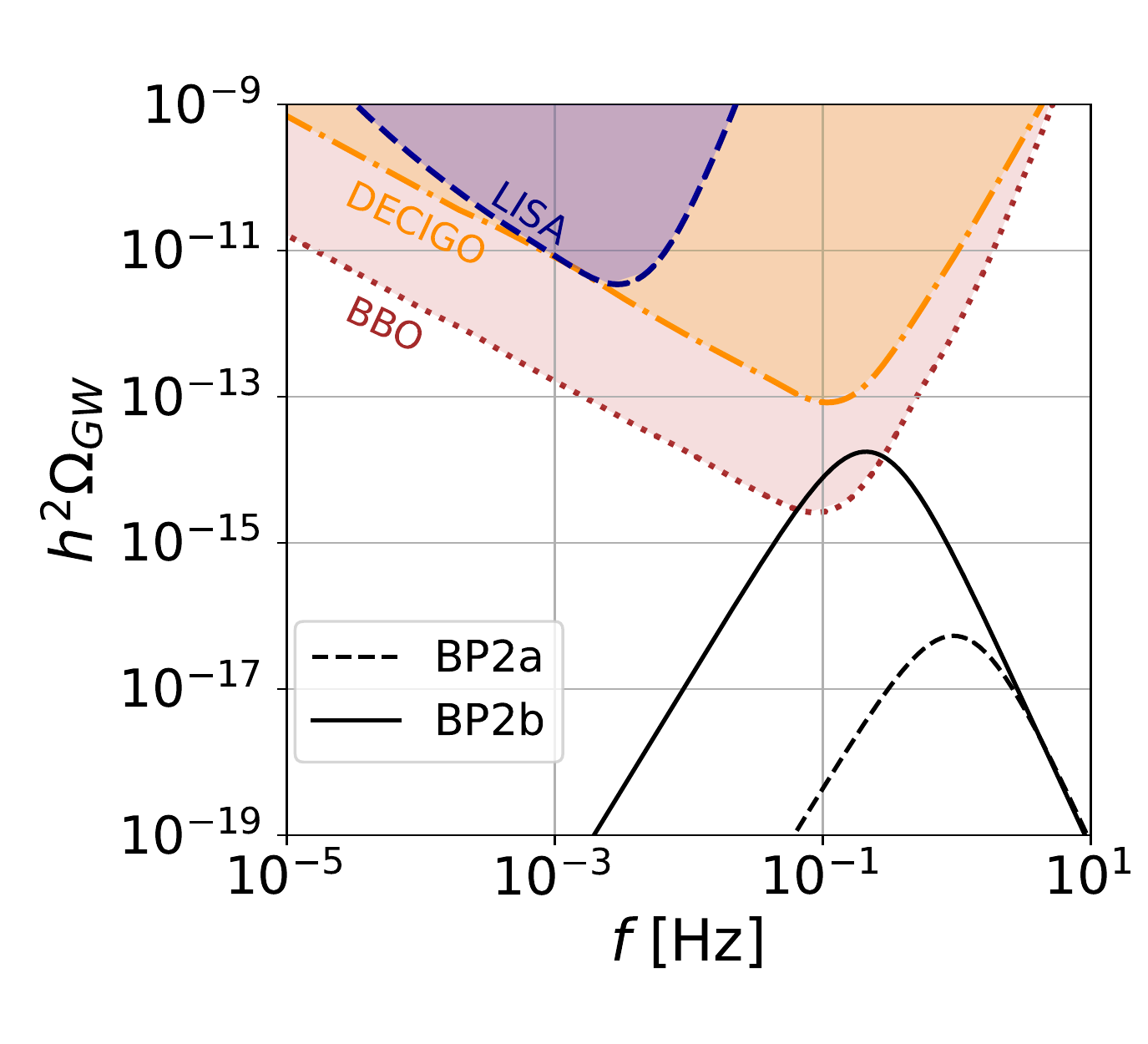}
    \caption{The GW spectra calculated at the benchmark points in Table~\ref{tab:BPs}. The purple, orange and red lines are the sensitivity curves of LISA, DECIGO, and BBO. The black solid and dashed lines are the GW spectra which are produced by the first order phase transition in these points.}
    \label{fig:GWspect}
\end{figure}
The GW spectra calculated at the benchmark points in Table~\ref{tab:BPs} are shown in figure~\ref{fig:GWspect}.
The purple, orange and red lines are the sensitivity curves of LISA~\cite{LISA:2017pwj}, DECIGO~\cite{Seto:2001qf}, and BBO~\cite{Corbin:2005ny} in ref.~\cite{Hashino:2018wee}.
The solid and dashed black lines in the left panel of figure~\ref{fig:GWspect} are the GW spectra at the BP1a and BP1b, respectively, while the ones at the BP2a and BP2b are shown in the right panel.
The BP1b whose velocity is 0.45 reaches the sensitivity curves of DECIGO and BBO, and the BP2b only reaches the one of BBO.
Therefore, the BP1b and BP2b can explain the observed BAU, and these also can be tested by the future space-based interferometers, in addition to collider signatures.

We next discuss testabilities of CP violation in the model at future experiments.
The parameter in our model $\zeta_d$, which controls the strength of down-type quark couplings to the additional Higgs bosons, can be constrained by future measurements of $B \to X_s \gamma$ and $\Delta \mathcal{A}_{CP}$ \cite{Modak_Senaha}.
The observable $\Delta \mathcal{A}_{CP}$ is related to CP violation in the process of $B \to X_s \gamma$.
It is defined by~\cite{Benzke:2010tq}
\begin{equation}
    \Delta \mathcal{A}_{CP} = A_{CP}(B^+ \to X_s^+ \gamma) - A_{CP}(B^0 \to X_s^0 \gamma),
\end{equation}
where,
\begin{align}
    &A_{CP}(B^+ \to X_s^+ \gamma) = \frac{\Gamma(B^- \to X_s^- \gamma) - \Gamma(B^+ \to X_s^+ \gamma)}{\Gamma(B^- \to X_s^- \gamma) + \Gamma(B^+ \to X_s^+ \gamma)}, \notag \\
    &A_{CP}(B^0 \to X_s^0 \gamma) = \frac{\Gamma(\overline{B}^0 \to \overline{X}_s^0 \gamma) - \Gamma(B^0 \to X_s^0 \gamma)}{\Gamma(\overline{B}^0 \to \overline{X}_s^0 \gamma) + \Gamma(B^0 \to X_s^0 \gamma)}.
\end{align}
By using the Wilson coefficients $C_7$ and $C_8$, $\Delta \mathcal{A}_{CP}$ is given by~\cite{Benzke:2010tq}
\begin{equation}
    \Delta \mathcal{A}_{CP} \simeq 0.12 \left( \frac{\tilde{\Lambda}_{78}}{100~\mathrm{MeV}} \right) \mathrm{Im}\left(\frac{C_8}{C_7}\right),
\end{equation}
where $\tilde{\Lambda}_{78}$ implies uncertainties from the hadronic scale, and it is estimated as $17 ~\mathrm{MeV} < \tilde{\Lambda}_{78} < 190~ \mathrm{MeV}$~\cite{Benzke:2010tq}.
In the following analysis, we set $\tilde{\Lambda}_{78} = 89$ MeV as the average value~\cite{Modak_Senaha}.
In the SM, $\Delta \mathcal{A}_{CP} = 0$ because both the Wilson coefficients $C_7$ and $C_8$ are real, so that it has a sensitivity to CP violation from new physics.
From the current experimental data at Belle~\cite{Belle:2018iff}, we obtain $\Delta \mathcal{A}_{CP} = (+3.69 \pm 2.65 \pm  0.76) \%$, where the first uncertainty is the statistical error and the second is the systematical one.
At Belle-II~\cite{Belle-II:2018jsg} with $50~\mathrm{ab}^{-1}$ as the future flavor experiment, it is expected that the absolute uncertainty is reduced to be 0.3 \%.
It is also expected at Belle-II with $50~\mathrm{ab}^{-1}$ that the relative uncertainty in the measurement of $B \to X_s \gamma$ can be reduced to be 3.2 \%~\cite{Belle-II:2018jsg}.

As we mentioned in section \ref{sec:CONST}, the nEDM can be used to constrain $\zeta_d$.
The upper bound of the nEDM is expected to be about one order higher accuracy in future nEDM experiments~\cite{Martin:2020lbx}.
However, there are still some uncertainties especially including the sign of the contribution from the Weinberg operator.

\begin{figure}[t]
    \centering
    \includegraphics[width=0.45\linewidth]{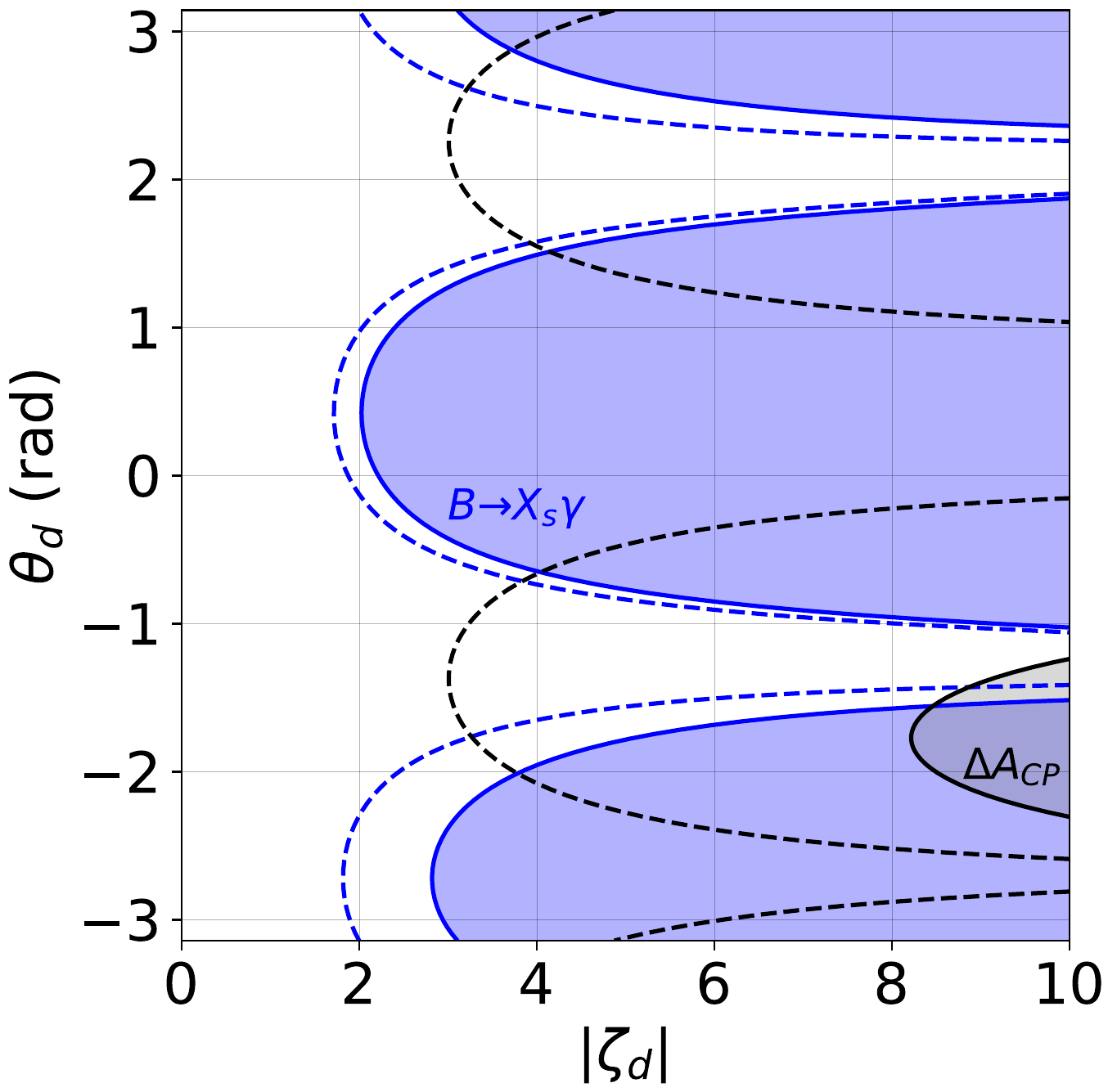}
    \includegraphics[width=0.45\linewidth]{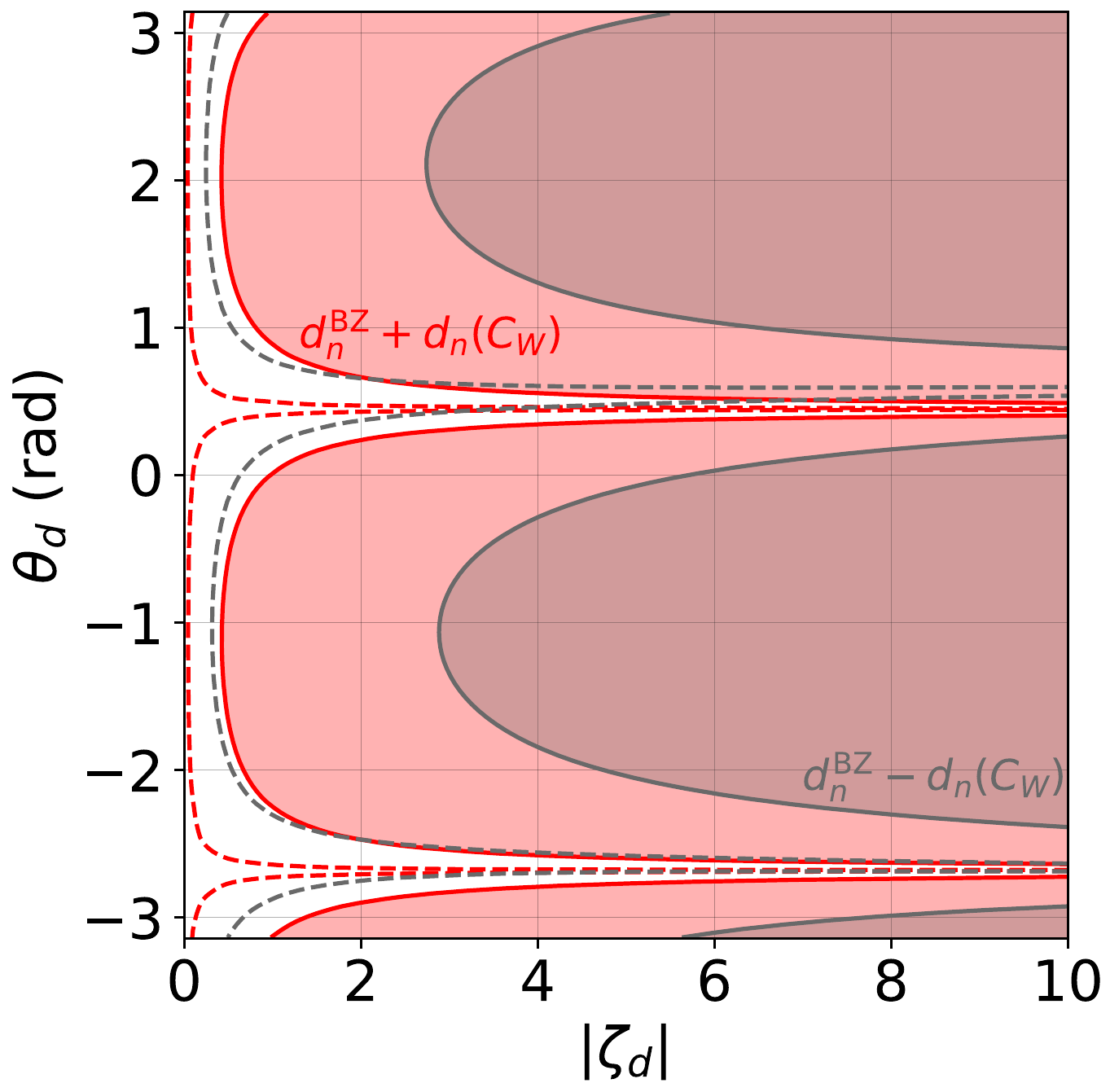}
    \caption{The constraints on $|\zeta_d|$-$\theta_d$ plane at the BP1 from the flavor (left) and nEDM (right) experiments. The blue (black) shaded regions in the left panel are the excluded regions from $B \to X_s \gamma$ ($\Delta \mathcal{A}_{CP}$). The red (gray) regions in the right panel are excluded by the current nEDM experimental data when the Weinberg operator positively (negatively) contributes to the nEDM. The dashed lines in each panel are future expected bounds.}
    \label{fig:zdconst}
\end{figure}
Figure~\ref{fig:zdconst} shows that the current and future expected bounds for $\zeta_d$.
The input parameters are the same as the BP1.
First, we explain the left panel of figure~\ref{fig:zdconst}.
Blue (black) regions are the excluded regions at $2 \sigma$ level from the current measurement of $B \to X_s \gamma$ ($\Delta \mathcal{A}_{CP}$), and the dashed lines are the future excluded bounds at Belle-II.
In the left figure, $B \to X_s \gamma$ and $\Delta \mathcal{A}_{CP}$ cover the different regions of $\theta_d$, and we find $|\zeta_d| \gtrsim 3$ can be excluded at Belle-II.

The nEDM constraints on $\zeta_d$ are shown in the right panel of figure~\ref{fig:zdconst}.
Red (gray) regions are excluded by the current experimental data~\cite{nEDM:2020crw} when the contribution from the Weinberg operator $d_n(C_W)$ positively (negatively) contributes.
The dashed lines are the one order higher accurate bound expected in future nEDM experiments~\cite{Martin:2020lbx}.
In the constructive case, where the total value of the nEDM is given by the sum of the two loop contribution $d_n^{\mathrm{BZ}}$ and the Weinberg operator contribution $d_n(C_W)$, the vast region of $\zeta_d$ has been already excluded except for $\theta_d \simeq \theta_u$ or $\theta_u + \pi$.
In such a case, the almost all regions would be excluded by future nEDM experiments.
On the other hand, in the destructive case, where the Weinberg operator negatively contributes to the nEDM, $|\zeta_d| \gtrsim 3$ has already been excluded at $\theta_d = \theta_u \pm \pi /2$, and $|\zeta_d| \gtrsim 0.3$ is expected to be excluded in the future.
As a result, $\zeta_d$ is constrained by the combination of $B \to X_s \gamma$ and $\Delta \mathcal{A}_{CP}$ at the Belle-II and the current nEDM constraint.
By future nEDM experiments, even in the destructive case, the large region of $\zeta_d$ can be constrained.

\section{Discussions\label{sec:DISCUSS}}
In this section, we give some comments on the results shown in section~\ref{sec:BAU}.
In section~\ref{sec:BAU}, we have considered the top transport scenario~\cite{Fromme:2006wx} in the aligned THDM, where the top quarks are the CP-violating source for the BAU.
However, the possibilities of EWBG where the light quarks or leptons become the CP-violating source were also discussed in refs.~\cite{Chung:2009cb, DeVries:2018aul, Xie:2020wzn, Chiang:2016vgf, Guo:2016ixx, Fuyuto_Senaha, Modak_Senaha}.
In our model, for example in the case of $|\zeta_e| \gg |\zeta_u|$, the tau leptons might be an important role of EWBG.
Furthermore, Yukawa couplings in our model can be generalized to be allowed flavor mixing structure.
This mixing is severely constrained from the flavor experiments, however in such a case, it is known that EWBG can be realized by using non-diagonal Yukawa couplings, e.g. the $t$-$c$ mixing~\cite{Fuyuto_Senaha} or the $\tau$-$\mu$ mixing~\cite{Chiang:2016vgf, Guo:2016ixx}.

In our numerical calculation for the BAU, we have used the WKB approximation method~\cite{Joyce:1994fu, Joyce:1994zn, Cline:2000nw, Fromme:2006wx, Cline:2020jre}.
There are other formalisms, which is so-called the VEV Insertion Approximation (VIA)~\cite{Riotto:1995hh, Riotto:1997vy}.
A difference of results for the BAU between WKB and VIA methods has been discussed in refs.~\cite{Basler:2021kgq, Cline:2020jre, Cline:2021dkf}.
The generated BAU in the VIA method tends to be orders of magnitude larger than that in the WKB method.
Therefore, if we use the VIA method, our benchmark points shown in section \ref{sec:BAU}, which can explain the observed BAU, might be changed.
Even in this case, we can explain the observed BAU by EWBG under the current experiments by reducing the parameters $|\zeta_u|$ and $|\lambda_7|$ with keeping the destructive interference of the eEDM.
Recently, in ref.~\cite{Postma:2022dbr}, it has been shown that the VIA source term which is obtained from Kadanoff--Baym equations at leading order in the derivative expansion exactly vanishes.
However, the origin of the different results between WKB and VIA methods is still unknown.

In order to calculate the wall velocity $v_w$, one has to solve equations of motion of the bubble in the fluid~\cite{Moore:1995ua, Moore:1995si}.
For simplicity, we have treated it as a free parameter.
The wall velocity $v_w$ has been calculated in several models~\cite{John:2000zq, Konstandin:2014zta, Kozaczuk:2015owa, Dorsch:2016nrg, Cline:2021iff}.
In the THDMs like our model, the order of magnitude of $v_w$ agrees with the SM one, and it is evaluated as $\mathcal{O} (10^{-1})$~\cite{Konstandin:2014zta, Dorsch:2016nrg}. 

For the successful EWBG, 
large scalar couplings are often necessary to realize the strongly first order EWPT in the aligned THDM. 
According to the discussion in ref.~\cite{Cline:2011mm}, 
in the case that the largest scalar coupling is $|\lambda| \sim 3$, 
the Landau pole is expected to appear at $1\text{--}3~\mathrm{TeV}$. 
On the other hand, in ref.~\cite{Dorsch:2016nrg}, the running couplings become non-perturbative at a scale higher than $4~\mathrm{TeV}$ at one-loop level even in the case that the largest coupling is $|\lambda| \sim 7$. 
This difference may be caused by the threshold effect of the running couplings.\footnote{
Another difference between refs.~\cite{Cline:2011mm} and~\cite{Dorsch:2016nrg} is the discrete symmetry of the Higgs potential. 
The $Z_2$ breaking terms $\lambda_6$ and $\lambda_7$ are not include in ref.~\cite{Dorsch:2016nrg} while they are included in ref.~\cite{Cline:2011mm}. 
However, the effect of this difference would be small because the relatively small couplings $\lambda_6 = 0$ and $|\lambda_7| \lesssim 0.6$ are considered in ref.~\cite{Cline:2011mm}.}
In ref.~\cite{Cline:2011mm}, the running effect of the additional Higgs bosons is included from the scale of $m_Z$. 
On the other hand, it is included above the scale of the mass of additional Higgs bosons in ref.~\cite{Dorsch:2016nrg}. 

In the case that the non-decoupling effect of the additional Higgs bosons is important like in our benchmark scenarios, 
it is not clear how we should handle the threshold effect of the additional Higgs bosons because the additional invariant mass scale $M$ is small and the additional Higgs bosons are not completely decoupled even in the low scale physics.  
The threshold effect would drastically change the Landau pole in the model as suggested by the difference between refs.~\cite{Cline:2011mm} and~\cite{Dorsch:2016nrg}. 
Consequently, a more detailed discussion is necessary to investigate the Landau pole in our benchmark scenario. 
In our paper, we have not analysed this issue, which will be given elsewhere~\cite{ref: future}. 

As shown in figure~\ref{fig:COLOR}, successful EWBG can be realized by the masses of the heavy Higgs bosons becoming around $300$-$400$ GeV.
In this mass region, $|\zeta_u|$ is constrained from above by $H_{2,3} \to \tau \tau$ and $H^\pm \to tb$ searches as shown in figure~\ref{fig:ConstDirect}.
Thus our model can be tested by the direct search of the heavy Higgs bosons at the HL-LHC.
If the additional scalar masses are smaller than about $300$ GeV, the multi lepton search at the HL-LHC might be used to test our model~\cite{Kanemura:2021dez}.
In our model, we have assumed the alignment condition $\lambda_6 = 0$ to avoid the mixing among the neutral scalar states.
If this alignment condition is slightly broken, decay branching ratios of the additional Higgs bosons, the vacuum stability, and the perturbative unitarity condition are changed.
As a result, the testability of the model at the HL-LHC and the future upgraded ILC can be much enhanced~\cite{Aiko:2020ksl, Aiko:2021can, Kanemura:2022ldq}.

The effect of the heavy Higgs bosons appear in the flavor physics, so that the future flavor experiments such as 
Belle-II~\cite{Belle-II:2018jsg} or LHCb~\cite{LHCb:2012myk} can be used to test the model.
As shown in figure \ref{fig:ConstDirect} and figure \ref{fig:zdconst}, observables of $B \to X_s \gamma$, $B_s \to \mu \mu$ and $\Delta \mathcal{A}_{\mathrm{CP}}$ have some sensitivities about the quantum loop effect of the heavy Higgs bosons or the CP violation in the model.

CP violation in the Higgs potential can be detected by the ILC and the measurements of the EDM.
As we have mentioned in section \ref{sec:BAU}, both the upper bounds of the eEDM and nEDM in future experiments have about an order higher accuracy than the current bounds~\cite{ACME:2018yjb, Martin:2020lbx}.
Therefore, for example by the future ACME experiment~\cite{ACME:2018yjb}, we can exclude many points in figure~\ref{fig:scat2}.
In the case of $|\zeta_e| \gg |\zeta_u|$ and $\theta_e = \mathcal{O} (1)$, the CP-violating phase of $\zeta_e$ in the model would be decided at the ILC by the measurement of the azimuthal angular distribution where a tau pair from decay of the additional neutral bosons decays into hadrons~\cite{Jeans:2018anq, Kanemura:2021atq}. 

In section~\ref{sec:BAU}, we have discussed the triple Higgs coupling, the GWs, and $H_1 \to \gamma \gamma$ as a probe of strongly first order phase transitions.
The triple Higgs coupling in our model is measured by the process of di-Higgs production at future colliders~\cite{Pairprod_had1, Pairprod_had2, Goncalves:2018qas, Tian:2013qmi, Kurata:2013, Fujii:2015jha, Asakawa:2010xj}.
At the HL- LHC and the ILC with $\sqrt{s} = 500$ GeV (1 TeV), this coupling is expected to be measured at the 50\% \cite{Cepeda:2019klc} and 27\% (10 \%)~\cite{Fujii:2015jha, Bambade:2019fyw} accuracy, respectively.
BP1 and BP2 in figure~\ref{fig:COLOR} and Table~\ref{tab:BPs}, $\Delta R = 41 \%$ and $66 \%$, respectively, so that the strongly first order phase transition in these benchmark points can be tested at these future colliders.

In Table~\ref{tab:BPs}, we have shown the branching ratio of $H_1 \to \gamma \gamma$ in the benchmark points.
In future collider experiments, the measurement of the Higgs di-photon decay would become more precise.
For example in the HL-LHC, the relative uncertainty of the branching ratio of $H_1 \to \gamma \gamma$ is expected to be 2.6\%~\cite{Cepeda:2019klc}.
Therefore, our model can also be tested via the precise measurement of the Higgs di-photon decay.

We have shown the GW spectra at some benchmark points in figure~\ref{fig:GWspect}, while these do not reach the sensitivity curve of LISA. 
Nevertheless, we expect that these GW spectra can be detected at LISA by using the Fisher matrix analysis discussed in ref.~\cite{Hashino:2018wee}.
The possibilities of detection of the GWs at DECIGO and BBO are also expected to be enhanced by using this analysis.
We can obtain a GW spectrum which has a larger height of the peak by being the phase transition stronger, however in such a case, the WKB approximation for the BAU is no longer valid because of decreasing $L_w T_n$.

\section{Conclusions\label{sec:CONCL}}

In this paper, We have discussed electroweak baryogenesis in the aligned THDM.
It has been known that in this model the dangerous constraint from the experiment for the eEDM can be avoided by the destructive interference among the CP-violating effects in the Higgs sector.
In our previous paper, we have shown that the observed baryon number of the Universe can be explained in a specific scenario in this model without contradicting current available data, and some phenomenological consequences are also discussed.
Here we have discussed details of the evaluation of baryon number based on the WKB method with taking into account all order of the wall velocity with all formulae.
We then have investigated parameter spaces which are allowed simultaneously under the current available data from collider, flavor and EDM experiments, and we have found several benchmark scenarios which can explain the BAU.
We have discussed how we can test these benchmark scenarios at future collider experiments, various flavor experiments and gravitational wave observations.
In particular, the model can be tested by the di-photon decay of the Higgs boson and the triple Higgs boson coupling due to the non-decoupling effect which causes strongly first order phase transition.
In the viable scenario with a relatively large wall velocity, enough amounts of gravitational waves are predicted for the observations at future space-based interferometers.

\section*{Acknowledgments}
The work of K. E. was supported in part by JSPS KAKENHI Grant No.~JP21J11444.
The work of S. K. was supported in part by JSPS KAKENHI Grants No.~20H00160 and No.~22F21324.
The work of Y. M. was supported by JST SPRING, Grant No.~JPMJSP2138.

\appendix
\section{\label{sec:APP}}

First, we show the explicit formulae of the field dependent mass included thermal corrections at one loop level.
We here denote the squared field dependent masses of the field $\phi$ as $\tilde{m}^2_{\phi \phi}$.
The matrix elements of the charged scalar states are given by
\begin{align}
    \tilde{m}_{G^+ G^-}^2 &= -\mu_1^2 + \frac{1}{2}\lambda_1 \varphi_1^2 + \frac{1}{2} \lambda_3 (\varphi_2^2 + \varphi_3^2) + \left(\lambda_{6R} \varphi_2 - \lambda_{6I} \varphi_3 \right)\varphi_1 
    \nonumber \\ &~~~+ \frac{T^2}{24} \left( 3\lambda_1 + 4\lambda_3 + 2\lambda_4 + 6y^2_t + \frac{9}{2}g^2 + \frac{3}{2}g^{\prime 2} \right), \nonumber \\
    \tilde{m}_{G^+ H^-}^2 &= -(\mu_3^2)^* + \frac{1}{2} \lambda_4 \varphi_1(\varphi_2 + i \varphi_3) + \frac{1}{2}\lambda_5^* \varphi_1(\varphi_2 - i \varphi_3) + \frac{1}{2}\lambda_6^* \varphi_1^2 + \frac{1}{2}\lambda_7^* (\varphi_2^2 + \varphi_3^2) \nonumber \\
    &~~~ + \frac{T^2}{24}(6 \lambda^*_6 + 6 \lambda_7^* + 6y^2_t\zeta_u^*), \nonumber \\
    \tilde{m}_{H^+ G^-}^2 &= (\tilde{m}_{G^+ H^-})^*, \nonumber \\
    \tilde{m}_{H^+ H^-}^2 &= -\mu_2^2 + \frac{1}{2}\lambda_2 (\varphi_2^2 + \varphi_3^2) + \frac{1}{2} \lambda_3 \varphi_1^2 + (\lambda_{7R}\varphi_2 - \lambda_{7I}\varphi_3)\varphi_1 \nonumber \\
    &~~~+\frac{T^2}{24} \left(6\lambda_2 + 4\lambda_3 + 2\lambda_4 + 6 y^2_t |\zeta_u|^2 + \frac{9}{2}g^2 + \frac{3}{2}g^{\prime 2} \right),
\end{align}
where $\star_{R}$ ($\star_{I}$) means real (imaginary) part, and $y_t$ is the top quark Yukawa coupling $\sqrt{2} m_t / v$.
The matrix elements of the neutral scalar states are given by
\begin{align}
    \tilde{m}_{\varphi_1 \varphi_1}^2 &= -\mu_1^2 + \frac{3}{2}\lambda_1 \varphi_1^2 + \frac{1}{2}(\lambda_3 + \lambda_4)(\varphi_2^2 + \varphi_3^2) + \frac{1}{2}\left( \lambda_{5R}(\varphi_2^2 - \varphi_3^2) - 2\lambda_{5I}\varphi_2\varphi_3 \right)  \nonumber \\
    &~~~ + 3(\lambda_{6R}\varphi_2 - \lambda_{6I}\varphi_3) \varphi_1 + \frac{T^2}{24}\left(6\lambda_1 + 4\lambda_3 + 2\lambda_4 + 6 y^2_t + \frac{9}{2}g^2 + \frac{3}{2}g^{\prime 2} \right), 
\end{align}
\begin{align}
    \tilde{m}_{\varphi_I \varphi_I}^2 &= -\mu_1^2 + \frac{1}{2}\lambda_1 \varphi_1^2 + \frac{1}{2}(\lambda_3 + \lambda_4)(\varphi_2^2 + \varphi_3^2) - \frac{1}{2}\left( \lambda_{5R}(\varphi_2^2 - \varphi_3^2) - 2\lambda_{5I}\varphi_2 \varphi_3 \right)  \nonumber \\
    &~~~ + (\lambda_{6R} \varphi_2 - \lambda_{6I}\varphi_3)\varphi_1 +\frac{T^2}{24} \left(6\lambda_1 + 4\lambda_3 + 2\lambda_4 + 6 y^2_t + \frac{9}{2}g^2 + \frac{3}{2}g^{\prime 2} \right), 
\end{align}
\begin{align}
    \tilde{m}_{\varphi_2 \varphi_2}^2 &= -\mu_2^2 + \frac{1}{2} \lambda_2 (3 \varphi_2^2 + \varphi_3^2) + \frac{1}{2}(\lambda_3 + \lambda_4)\varphi_1^2 + \frac{1}{2}\lambda_{5R}\varphi_1^2 + (3\lambda_{7R}\varphi_2 - \lambda_{7I}\varphi_3)\varphi_1 \nonumber \\
    &~~~ +\frac{T^2}{24}\left(6\lambda_2 + 4\lambda_3 + 2\lambda_4 + 6y^2_t|\zeta_u|^2 + \frac{9}{2}g^2 + \frac{3}{2}g^{\prime 2} \right), 
\end{align}
\begin{align}
    \tilde{m}_{\varphi_3 \varphi_3}^2 &= -\mu_2^2 + \frac{1}{2} \lambda_2 (\varphi_2^2 + 3 \varphi_3^2) + \frac{1}{2}(\lambda_3 + \lambda_4)\varphi_1^2 - \frac{1}{2}\lambda_{5R}\varphi_1^2 + (\lambda_{7R}\varphi_2 - 3\lambda_{7I}\varphi_3)\varphi_1 \nonumber \\
    &~~~ +\frac{T^2}{24}\left(6\lambda_2 + 4\lambda_3 + 2\lambda_4 + 6y^2_t|\zeta_u|^2 + \frac{9}{2}g^2 + \frac{3}{2}g^{\prime 2} \right), 
\end{align}
\begin{align}
    \tilde{m}_{\varphi_1 \varphi_I}^2 &= \frac{1}{2}\left(2\lambda_{5R} \varphi_2 \varphi_3 + \lambda_{5I}(\varphi_2^2 - \varphi_3^2) \right) + (\lambda_{6I} \varphi_2 + \lambda_{6R}\varphi_3)\varphi_1,\qquad\qquad\qquad\qquad 
\end{align}
\begin{align}
    \tilde{m}_{\varphi_1 \varphi_2}^2 &= -\mu_{3R}^2 + (\lambda_3 +\lambda_4)\varphi_1 \varphi_2 + (\lambda_{5R} \varphi_2 - \lambda_{5I}\varphi_3) \varphi_1 + \frac{3}{2}\lambda_{6R}\varphi_1^2  \nonumber \\
    &~~~ + \frac{1}{2}(3\lambda_{7R} \varphi_2^2 + \lambda_{7R} \varphi_3^2 - 2\lambda_{7I}\varphi_2\varphi_3) +\frac{T^2}{24}(6\lambda_{6R} + 6 \lambda_{7R} + 6y^2_t|\zeta_u|\cos{\theta_u}),
\end{align}
\begin{align}
    \tilde{m}_{\varphi_1 \varphi_3}^2 &= \mu_{3I}^2 + (\lambda_3 +\lambda_4)\varphi_1 \varphi_2 - (\lambda_{5I} \varphi_2 + \lambda_{5R}\varphi_3) \varphi_1 - \frac{3}{2}\lambda_{6I}\varphi_1^2  \nonumber \\
    &~~~ - \frac{1}{2}(\lambda_{7I} \varphi_2^2 + 3\lambda_{7I} \varphi_3^2 - 2\lambda_{7R}\varphi_2\varphi_3) +\frac{T^2}{24}(-6\lambda_{6I} - 6 \lambda_{7I} - 6y^2_t|\zeta_u|\sin{\theta_u}), 
\end{align}
\begin{align}
    \tilde{m}_{\varphi_I \varphi_2}^2 &= -\mu_{3I}^2 + (\lambda_{5I} \varphi_2 + \lambda_{5R}\varphi_3) \varphi_1 + \frac{1}{2}\lambda_{6I}\varphi_1^2 + \frac{1}{2}(3\lambda_{7I} \varphi_2^2 + \lambda_{7I} \varphi_3^2 + 2\lambda_{7R}\varphi_2\varphi_3) \nonumber \\
    &~~~ +\frac{T^2}{24}(6\lambda_{6I} + 6 \lambda_{7I} + 6y^2_t|\zeta_u|\sin{\theta_u}), 
\end{align}
\begin{align}
    \tilde{m}_{\varphi_I \varphi_3}^2 &= -\mu_{3R}^2 + (\lambda_{5R} \varphi_2 - \lambda_{5I}\varphi_3) \varphi_1 + \frac{1}{2}\lambda_{6R}\varphi_1^2 + \frac{1}{2}(\lambda_{7R} \varphi_2^2 + 3\lambda_{7R} \varphi_3^2 + 2\lambda_{7I}\varphi_2\varphi_3) \nonumber \\
    &~~~ +\frac{T^2}{24}(6\lambda_{6R} + 6 \lambda_{7R} + 6y^2_t|\zeta_u|\cos{\theta_u}), 
\end{align}
\begin{align}
    \tilde{m}_{\varphi_2 \varphi_3}^2 &= \lambda_2 \varphi_2 \varphi_3 - \frac{1}{2}\lambda_{5I} \varphi_1^2 + (\lambda_{7R} \varphi_3 - \lambda_{7I} \varphi_2) \varphi_1.\qquad\qquad\qquad\qquad\qquad\qquad
\end{align}
The symbol $\varphi_I$ is the imaginary part of the neutral component of $\sqrt{2} \langle \Phi_1 \rangle$, and it can be set to zero by the $SU(2)_L$ transformation.
The field dependent masses of the gauge bosons are given by
\begin{align}
&\tilde{m}_{W^a_\mu W^b_\nu}^2 = \left(\frac{1}{4}g^2(\varphi_1^2 + \varphi_2^2 + \varphi_3^2) + 2g^2 T^2 \delta_{\mu \mathrm{L}} \right)\delta_{ab}\delta_{\mu \nu},  \\
&\tilde{m}_{B_\mu B_\nu}^2 = \left( \frac{1}{4}g^{\prime 2}(\varphi_1^2 + \varphi_2^2 + \varphi_3^2) + 2g^{\prime 2} T^2 \delta_{\mu \mathrm{L}}  \right)\delta_{\mu \nu},  \\
&\tilde{m}_{W^a_\mu B_\nu}^2 = -\frac{1}{4}gg^\prime (\varphi_1^2 + \varphi_2^2 + \varphi_3^2) \delta_{\mu \nu},
\end{align}
where the thermal corrections only contribute to the longitudinal mode ($\mathrm{L}$) of the gauge bosons.
The top quark mass is given by
\begin{equation}
    \tilde{m}_t^2 = \frac{1}{2}y_t^2 \Big\{ \big(\varphi_1 + |\zeta_u|(\cos\theta_u \varphi_2 - \sin\theta_u \varphi_3) \big)^2 + |\zeta_u|^2 (\sin\theta_u \varphi_2 - \cos\theta_u \varphi_3)^2 \Big\}.
\end{equation} 

Next, the counter terms of the potential $V_{CT}$ are given by
\begin{align}
    V_{CT} = &-\frac{1}{2} \delta \mu_1^2 \varphi_1^2  - (\delta \mu_{3R}^2 \varphi_2 - \delta \mu_{3I}^2 \varphi_3)\varphi_1 \nonumber \\
    &+\frac{1}{8}\delta \lambda_1 \varphi_1^4 + \frac{1}{4} \delta \lambda_3 (\varphi_2^2 + \varphi_3^2)\varphi_1^2 \nonumber \\
    &+\frac{1}{4}\delta \lambda_{5R}(\varphi_2^2 - \varphi_3^2)\varphi_1^2 - \frac{1}{2} \delta \lambda_{5I}\varphi_1^2 \varphi_2 \varphi_3 + \frac{1}{2} \delta \lambda_{6R}\varphi_1^3 \varphi_2 - \frac{1}{2}\delta \lambda_{6I} \varphi_1^3 \varphi_3,
\end{align}
where each coupling is determined by the renormalization conditions in Eqs.~(\ref{eq: renormalization_cond_1}) and (\ref{eq: renormalization_cond_2});
\begin{align}
&\delta \mu_1^2 = \frac{1}{2}\left(\frac{3}{v} \frac{\partial V_1}{\partial \varphi_1} - \frac{\partial^2 V_1}{ \partial \varphi_1^2}\right), & & \notag \\
&\delta \mu_{3R}^2 = \frac{1}{2}\left(\frac{3}{v}\frac{\partial V_1}{\partial \varphi_2} - \frac{\partial^2 V_1}{\partial \varphi_1 \partial \varphi_2} \right), & &\delta \mu_{3I}^2 = \frac{1}{2} \left(-\frac{3}{v} \frac{\partial V_1}{\partial \varphi_3} + \frac{\partial^2 V_1}{\partial \varphi_1 \partial \varphi_3}\right), \notag \\
&\delta \lambda_1 v^2 = \frac{1}{v}\frac{\partial V_{1}}{\partial \varphi_1} - \frac{\partial^2 V_1}{\partial \varphi_1^2}, & &\delta \lambda_{3}v^2 = - \frac{\partial^2 V_1}{\partial \varphi_2^2} - \frac{\partial^2 V_1}{\partial \varphi_3^2}, \notag \\
&\delta \lambda_{5R} v^2 = - \frac{\partial^2 V_1}{\partial \varphi_2^2} + \frac{\partial^2 V_1}{\partial \varphi_3^2} , & &\delta \lambda_{5I} v^2 = 2\frac{\partial^2 V_1}{\partial \varphi_2 \partial \varphi_3}, \notag \\
&\delta \lambda_{6R} v^2 = \frac{1}{v}\frac{\partial V_1}{\partial \varphi_2} - \frac{\partial^2 V_1}{\partial \varphi_1 \partial \varphi_2}, & &\delta \lambda_{6I} v^2 = -\frac{1}{v} \frac{\partial V_1}{\partial \varphi_3} + \frac{\partial^2 V_1}{\partial \varphi_1 \partial \varphi_3}.
\end{align}
These results are consistent with ref.~\cite{Cline:2011mm}.

\end{document}